%% file: HAraujo.tex
\documentclass{elsart}
\usepackage{epsfig, amssymb, rotating}

\begin{document}

\begin{frontmatter}

\title{Detailed Calculation of Test-Mass Charging in the LISA Mission}
\author{H. M. Ara\'ujo,  P. Wass, D. Shaul, G. Rochester and T. J. Sumner}
\address{Blackett Laboratory, Imperial College London, SW7 2BW, UK}

\begin{abstract}
The electrostatic charging of the LISA test masses due to exposure of
the spacecraft to energetic particles in the space environment has
implications in the design and operation of the gravitational inertial
sensors and can affect the quality of the science data. Robust
predictions of charging rates and associated stochastic fluctuations
are therefore required for the exposure scenarios expected throughout
the mission. We report on detailed charging simulations with the
Geant4 toolkit, using comprehensive geometry and physics models, for
Galactic cosmic-ray protons and helium nuclei. These predict positive
charging rates of 50~+e/s (elementary charges per second) for solar
minimum conditions, decreasing by half at solar maximum, and current
fluctuations of up to 30~+e/s/Hz$^{1/2}$. Charging from sporadic solar
events involving energetic protons was also investigated. Using an
event-size distribution model, we conclude that their impact on the
LISA science data is manageable. Several physical processes hitherto
unexplored as potential charging mechanisms have also been
assessed. Significantly, the kinetic emission of very low-energy
secondary electrons due to bombardment of the inertial sensors by
primary cosmic rays and their secondaries can produce charging
currents comparable with the Monte Carlo rates.
\end{abstract}


\end{frontmatter}


\section{Introduction}
\label{intro}

LISA, the Laser Interferometer Space Antenna, is a joint ESA/NASA
mission designed to detect gravitational waves by means of space-borne
interferometric measurements. Gravitational disturbances will induce
minute variations in the relative separation of free-floating test
masses (TMs), which are distributed in pairs in an equilateral
constellation of three spacecraft 5 million km apart \cite{prephaseA}.

The radiation environment which will be encountered by LISA in its
Earth-like orbit around the Sun can threaten its ambitious target
sensitivity for detecting gravitational waves. The precursor mission
LISA Pathfinder (formerly known as SMART-2) will be similarly
vulnerable at the Earth-Sun L1 Lagrange point
\cite{vitale02}. Spurious TM accelerations result from the
accumulation of charge on the isolated masses caused by the
bombardment of the spacecraft by cosmic rays and solar particles. A
charged TM interacts electromagnetically with surrounding conducting
surfaces as well as with magnetic fields in interplanetary
space. These spurious Coulomb and Lorentz forces lead to both coherent
signals and in-band noise which can limit the sensitivity to
gravitational waves.

TM charging is a consideration in the design of the inertial sensor
(electromagnetic analysis, materials, shielding). Subsequently, it
drives the development of an effective charge management system
(charge measurement, neutralisation, operation modes, recovery from
drastic charging events). The unavoidable contamination of the science
data needs to be assessed and mitigated using data analysis
techniques, possibly assisted by information from diagnostic tools
such as particle monitors.

Previous work has established approximate charging rates from Galactic
cosmic rays at solar minimum conditions and the magnitude of their
spurious acceleration signals
\cite{jafry96,jafry97,sumner00,araujo03,shaul04a,shaul04b}. These
studies were based on Monte Carlo (MC) simulations using the Geant3
\cite{geant3} and Geant4 \cite{geant4} toolkits. Simulation work with
the Fluka package \cite{fluka} is also ongoing \cite{vocca04}.
Positive charging rates in the range 11--29~+e/s result from these
studies,\footnote{Note: a normalisation error was made in
Ref.~\cite{araujo03} in scaling the MC to the omnidirectional GCR
flux. Consequently, charging rate estimates found there are inflated
by a factor of 2. This has been accounted for in the comparisons
presented in this article.} depending on the complexity of the models
and TM size considered. In this work we set out to improve on previous
calculations for LISA by using more sophisticated physics models, a
detailed implementation of the spacecraft geometry, and by assessing
other possible charging mechanisms not considered previously. A
similar study for LISA Pathfinder is described elsewhere
\cite{wass04}.

\section{Geant4 Simulation}

\subsection{Geometry Model}

\begin{figure}[ht]
   \centerline{\epsfig{file=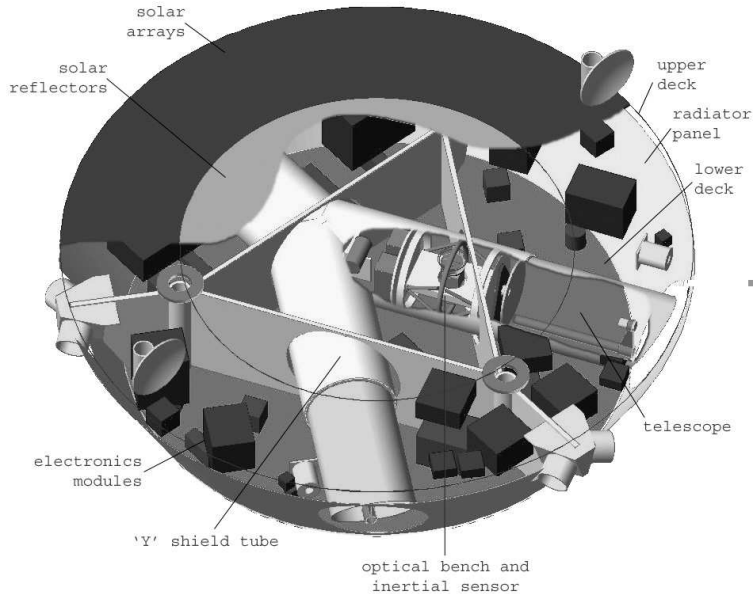,width=10cm,clip=}}
   \caption{LISA spacecraft model implemented in Geant4. Two inertial
    sensors are located inside the Y-shaped payload structure (one is
    shown in the figure).}
   \label{LISASpacecraft}
\end{figure}

The Geant4 implementation of the LISA Science Module, illustrated in
Fig.~\ref{LISASpacecraft}, was based on the LISA Integrated Solid
Model \cite{LISASolidModel}. Some 200 volumes represent nearly all
components above $\sim$0.1~kg at the correct location, using the
assigned materials when known. Overall, almost 85\% of a total
spacecraft mass of 400~kg has been accounted for; numerous small
pieces of structural hardware make up most of the missing mass.

\begin{figure}[ht]
   \centerline{\epsfig{file=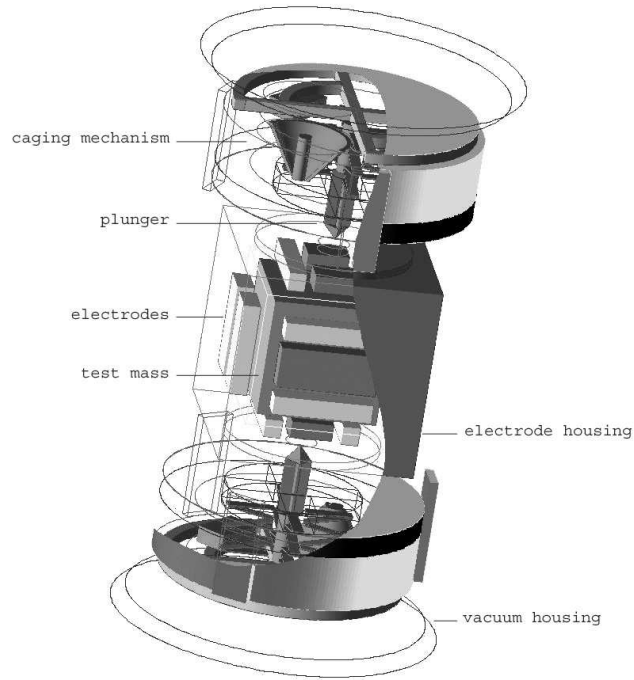,width=10cm,clip=}}
   \caption{LTP inertial sensor implemented in Geant4. The TM, located
   at the centre of the figure, is surrounded by sensing electrodes
   (light grey) and injection electrodes (dark grey). Caging
   mechanisms can be seen above and below the electrode housing.}
   \label{LISAInertialSensor}
\end{figure}

Each LISA module accommodates two interferometer telescopes mounted
inside a Y-shaped payload structure. The outer spacecraft structure is
approximately 2.7~m in diameter and 0.5~m in height. Two TMs are
housed in inertial sensors (IS) located in optical benches mounted
behind the telescopes. The IS representation is based on the current
design of the LISA Technology Package (LTP) sensor
~\cite{ISSolidModel} aboard the LISA Pathfinder. An important
difference between the earlier LISA and the new LTP sensor designs is
a larger TM considered in the latter (46~mm cube, as opposed to
40~mm). The Geant4 model is represented in a cutaway view in
Fig.~\ref{LISAInertialSensor}. The test cube is modelled as
Au$_{.7}$Pt$_{.3}$ alloy and is surrounded by sensing and actuation
electrodes ($Y$-$Z$ injection configuration) lodged in a molybdenum
housing. A 0.3~$\mu$m gold layer plates the entire inner surface of
the sensor housing. Caging mechanisms, used to immobilise the cube
during launch, are shown in Fig.~\ref{LISAInertialSensor} above and
below the sensor. The assembly is accommodated in a titanium vacuum
enclosure, the edges of which are also shown.

\subsection{Simulation of the Radiation Environment}

In the science phase of the LISA mission, two particle fluxes will
dominate the charging rates occurring in its test masses: a permanent
background flux of Galactic cosmic-ray (GCR) protons and light nuclei
and, occasionally, solar energetic particles (SEP) driven by shock
acceleration from certain types of solar event.

The GCR spectrum varies in intensity during the 11-year long solar
cycle. The interplanetary magnetic field is weakest at the minimum of
solar activity; as a consequence, a higher GCR flux is expected in the
solar cavity. At solar quiet, approximately 90\% of the particle flux
consists of protons, 8\% of He nuclei ($^3$He and $^4$He), 1\% of
heavier nuclei and 1\% of electrons. Conversely, the lowest GCR fluxes
are expected at solar maximum. The three most abundant primary nuclei
(p, $^3$He, $^4$He) at solar minimum and maximum fluxes were the main
simulation inputs. The TM charging from SEP events was determined
separately, and is discussed in Section~\ref{SEP}. The effect of
heavier nuclei is being examined, but it is not thought to be a
significant one. Work is under way to address the issue of GCR
variability in the LISA bandwidth and flux directionality. In the
present work GCR fluxes are considered constant and isotropic, which
should not affect the main conclusions reached.

\begin{figure}[ht]
   \centerline{\epsfig{file=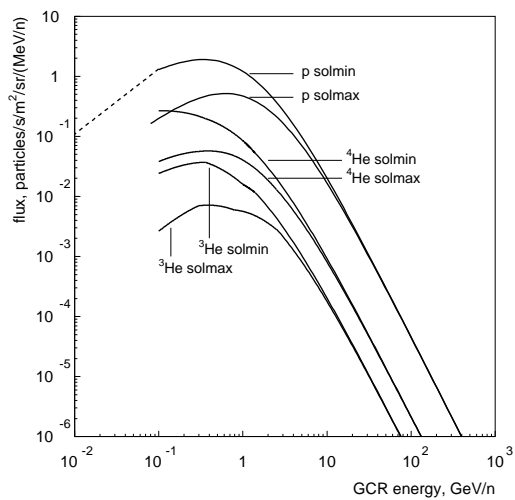,width=7cm}}
   \caption{Differential energy spectra for GCR protons and He nuclei
   at 1~AU.}
   \label{primaries}
\end{figure}

The adopted GCR spectra are those shown in Fig.~\ref{primaries}
\cite{grimani03}. Since most spectral data do not distinguish between
the two He isotopes, a simple parameterisation of the ratio
$^3$He/$^4$He was used \cite{catia}. Only energies above 100~MeV/n
were considered for the simulation input, the minimum required to
penetrate the shielding overlying the TMs, except for the solar
minimum proton spectrum, which was extended down to 10~MeV/n in order
to establish this fact.

In carrying out the MC simulation, the isotropic primary flux is
generated from points sampled uniformly from a spherical surface
encompassing the whole geometry, biased with a cosine-law angular
distribution around the surface normal. The particle energy is sampled
randomly from the GCR spectra. The MC timelines are normalised by
considering that $N_0$ primaries represent a fluence $\Phi=N_0/(\pi
R^2)$ particles per unit area anywhere inside a generator surface of
radius $R$.

\subsection{Physics Models}
\label{physics}

Owing to their high energy and hadronic nature, GCR interactions
entail complex nuclear reactions which have large final-state
multiplicities, producing a plethora of secondaries. All these
particles need to be tracked down to the lowest possible energy,
especially those approaching the IS. As a result, many of the physics
models available in Geant4 are required for this simulation; they
describe hadronic, electromagnetic and photonuclear interactions,
spanning almost 10 orders of magnitude in energy. A detailed
description of these models can be found in Ref.~\cite{physicsman}.

A milestone in the application of Geant4 to cosmic-ray simulations was
the implementation of intra-nuclear cascade models for the inelastic
scattering of nucleons, followed by their recent extension to light
nuclei. For protons, these are complemented by a quark-gluon string
model above $\sim$6~GeV. Nuclear evaporation models treat the excited
residual nucleus as well as primaries below $\sim$100~MeV. The new
models improve the treatment of GCR protons and allow, for the first
time, a realistic simulation of He fluxes.

A threshold of 250~eV was imposed for secondary particle production
inside the IS, with the exception of delta rays (knock-on electrons)
directly created by hadron ionisation; these are currently suppressed
below the mean excitation energy of the material ($\sim$800~eV in
gold). The effect of these production cuts is analysed later. Another
important issue in this simulation is the backscattering of electrons
at material boundaries. Backscattering fractions for heavy elements
such as gold and platinum can be as large as 50\% above keV
energies. The Geant4 model was tuned to reproduce the fraction, energy
and angular distributions of backscattered electrons found
experimentally for these materials.

The charging potential of several additional physics processes was
assessed separately from the simulation. Mechanisms such as the
kinetic emission of very low energy electrons and atoms (sputtering)
are not modelled by Geant4. Others, such as transition radiation and
Cherenkov emission, could have been included in the MC, but we opted
for a separate treatment due to their intricate dependence on the
detailed IS design. A prototype model of proton-induced x-ray emission
does exist in Geant4, but it proved too inaccurate for our
purpose. Finally, mainly for a speedier simulation, the effects of
cosmogenic activation and radioactive decay of the spacecraft
materials were investigated independently. All these processes are
discussed in Section~\ref{nonsim}.

\section{Simulation Results}

The MC model tracks cosmic rays and secondary particles capable of
penetrating the IS shielding (and all secondaries produced inside it),
tallying those entering and leaving the two test masses (TM$_0$ and
TM$_1$). Generation of the MC data required $\sim$1~CPU year to
achieve convergence of the charging rates to within a few percent
\cite{LSF}. A minimum exposure time $T$=400~s was simulated for both
solar minimum and solar maximum conditions, totalling $N_0=2.2\times
10^8$ simulated events, distributed as shown in
Table~\ref{table1}. The omnidirectional GCR fluxes above 100~MeV, $F$,
are also indicated. On average, one charging event was registered for
$\sim$1000 primaries striking the spacecraft. Some events were found
to deposit charge on both masses simultaneously, but their number was
negligible.

\input{table1.tex}

\subsection{Mean Charging Rates and Fluctuations}

The average charging rate of each TM, $R\pm\sigma_M$, and the spectral
density $S_R$ of the shot noise associated with the charging current
are summarised in Table~\ref{table1}, for a constant and isotropic GCR
flux. The MC uncertainty, $\sigma_M$, is calculated by combining the
Poisson variances for the occurrence of each net event charge. All
contributions to the total charging rate are positive for both solar
conditions, and slightly different for the two masses (but almost
within errors). A small discrepancy is to be expected: although the IS
occupy almost symmetrical positions in the spacecraft, they were
placed with the same orientation inside the payload arms and not as
mirror images with respect to the axis of symmetry. As a result, the
masses see slightly different amounts of shielding. For TM$_1$,
located in the IS visible in Fig.~\ref{LISASpacecraft}, the rates are
23.7$\pm$0.6~+e/s and 49.9$\pm$0.9~+e/s for solar maximum and solar
minimum fluxes, respectively. The solar minimum value is almost twice
the corrected figure obtained from Ref.~\cite{araujo03}
(29~+e/s). Although the proton flux clearly dominates these rates, it
is worth noting that He, which constitutes only 8\% of the total GCR
flux, contributes about twice that fraction to the total charging
rate.

\begin{figure}[ht]
  \centerline{
    \epsfig{file=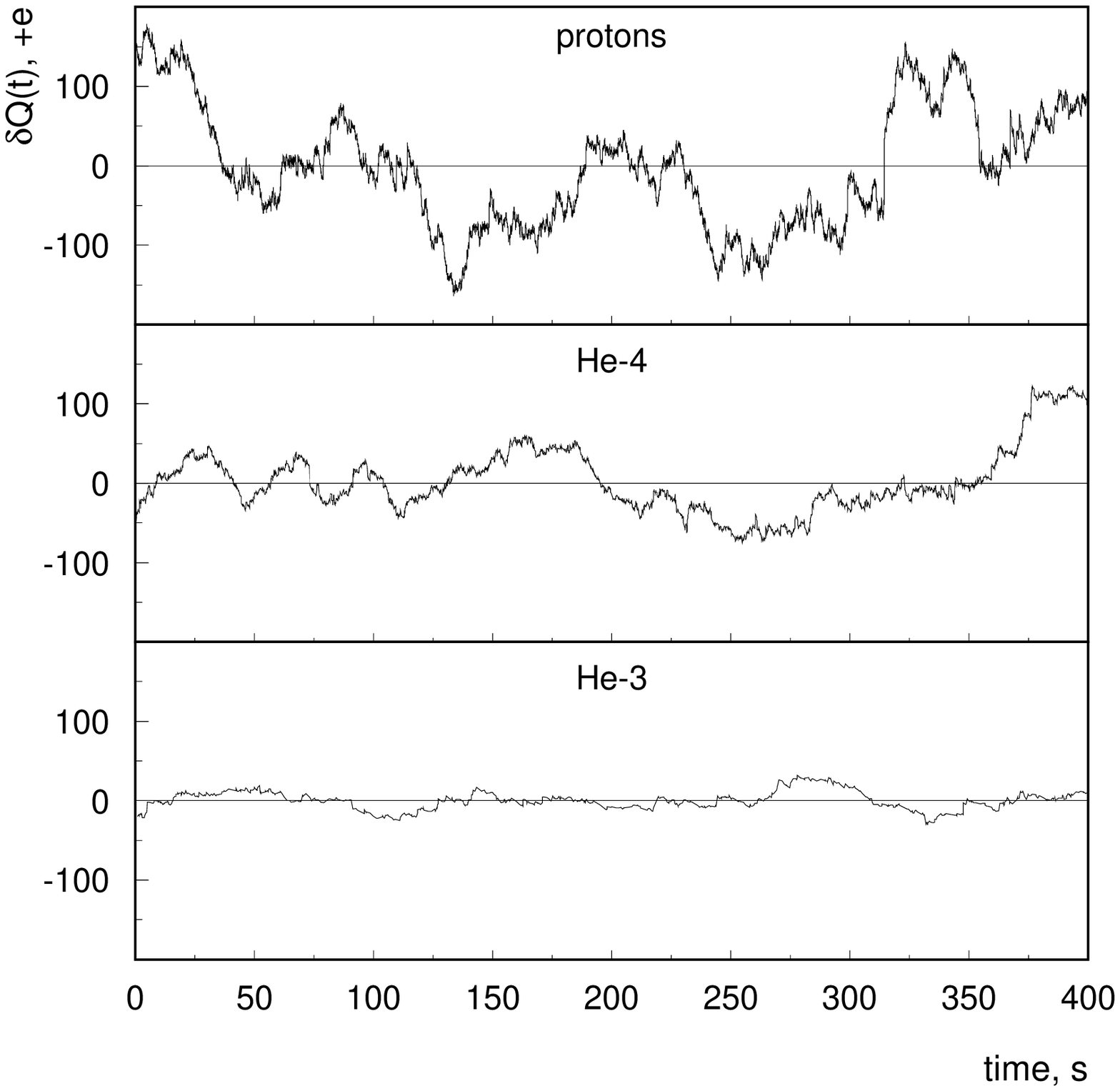,width=7cm}}
   \caption{Statistical fluctuations of the time-dependent net charge
   on TM$_0$ around the average net charging rate for solar minimum
   conditions.}
   \label{noise}
\end{figure}
To highlight the stochastic nature of the charging process, we write
the time-dependent fluctuations of the total charge as $\delta Q(t) =
Q(t) - R\,t$, shown in Fig.~\ref{noise}. Besides adding to the
acceleration noise, these fluctuations also set the ultimate accuracy
expected from a finite measurement of the total TM charge.

\begin{figure}[ht]
  \centerline{
    \epsfig{file=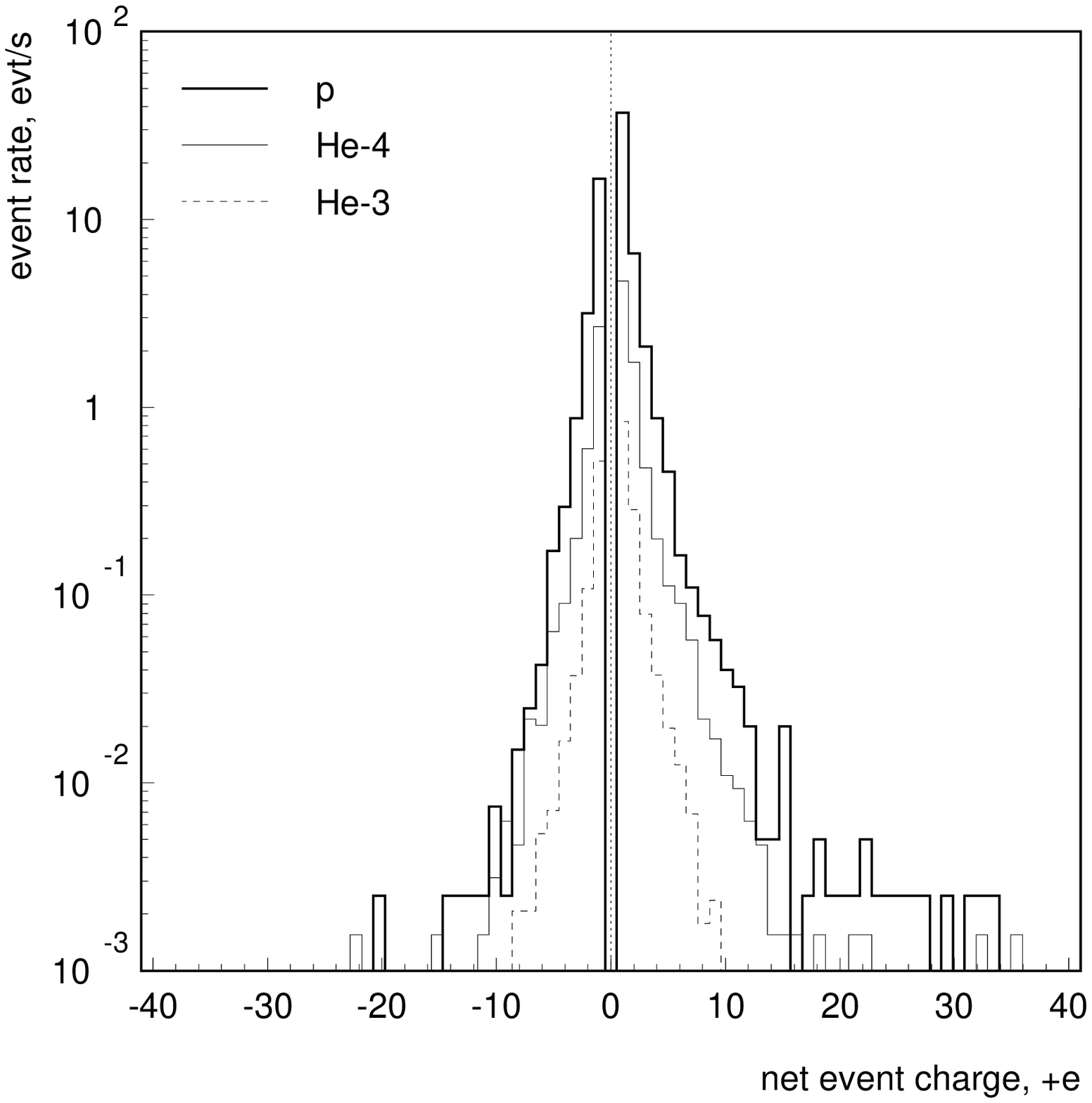,width=7cm}}
   \caption{Charge spectrogram for TM$_1$ at solar minimum.}
   \label{charge}
\end{figure}

For a more rigorous analysis, we write $R$ in terms of statistically
independent contributions from events giving only a net charge $qe$
($q$=+1 for protons). The parallel can then be made with the transport
of equal charges to and from a charge reservoir, i.e. electrical shot
noise. The distribution of individual charging currents is represented
in the spectrogram shown in Fig.~\ref{charge}. Although there is a
degree of balance between negative and positive net charges, the
latter dominate to produce a net positive rate. A few events were
recorded with even larger net charges than Fig.~\ref{charge} may
suggest. The largest found in the 6 datasets was +80~+e, resulting
from a 7~GeV proton; in another event, over 200 electrons entered one
TM leaving a net charge of -20~+e.

The shot noise for a particle rate $R_q/qe$ has a single-sided
spectral density equal to $S_q=(2(qe)R_q)^{1/2}$
[+e/s/Hz$^{1/2}$]. $S_R$ is obtained by adding $S_q$ in quadrature
over all values of $q$. The result is 24~+e/s/Hz$^{1/2}$ and
20~+e/s/Hz$^{1/2}$ for solar minimum and solar maximum,
respectively. Finally, the charge fluctuations at frequency $f$ are
obtained by integrating $S_R$ in the time domain: $S_Q(f)=S_R/2\pi f$.

The charging noise can also be described by an `effective rate' of
single charges required to produce the same spectral density,
$R_{e\!f\!f} = S_R^2/2$ . The solar maximum and minimum effective
rates are 196~+e/s and 288~+e/s, respectively. The latter figure is
also twice that previously calculated in Ref.~\cite{araujo03}.

\subsection{Charging as a Function of GCR Energy}

\begin{figure}[ht]
  \centerline{
    \epsfig{file=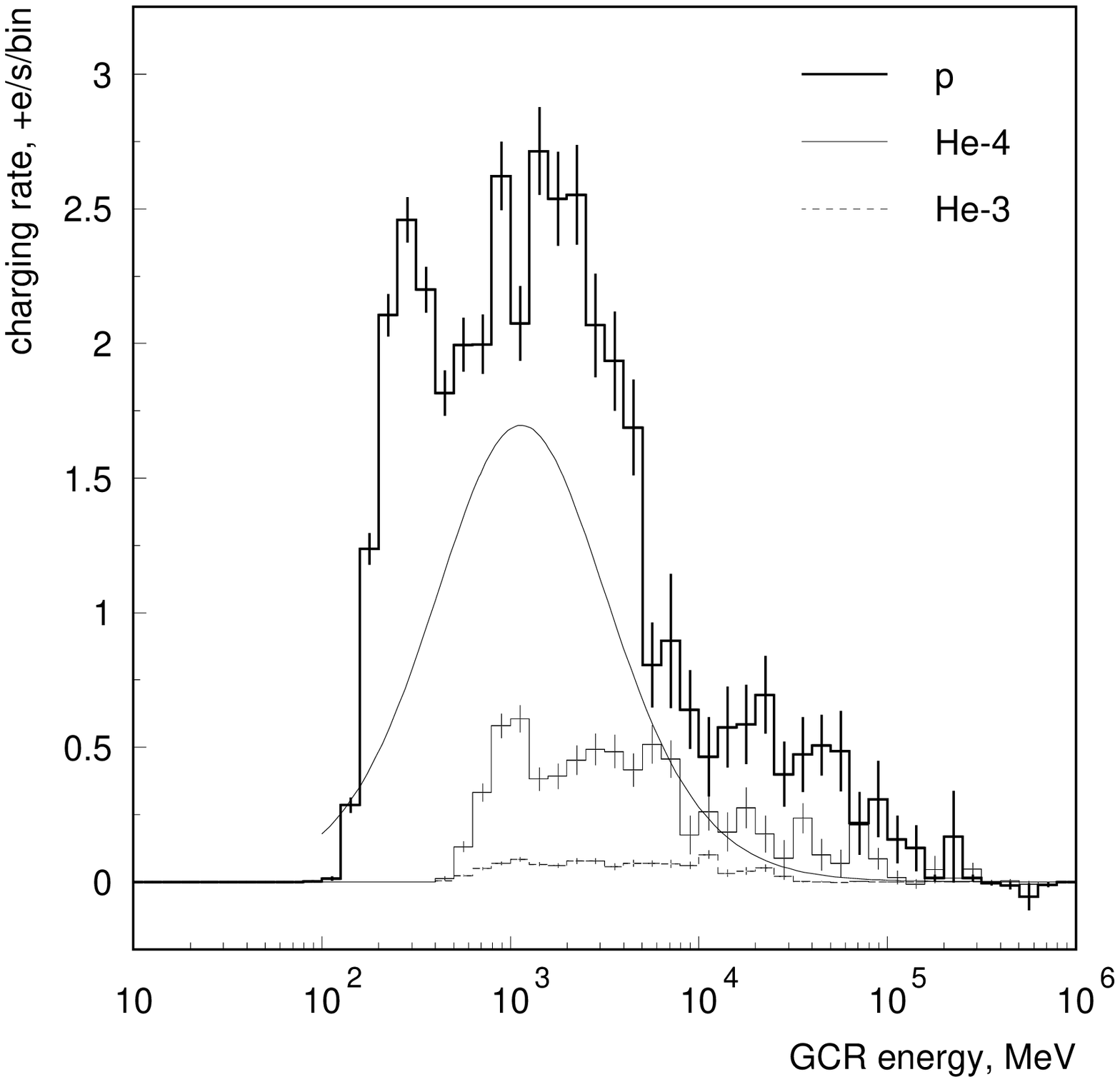,width=7cm}
    \epsfig{file=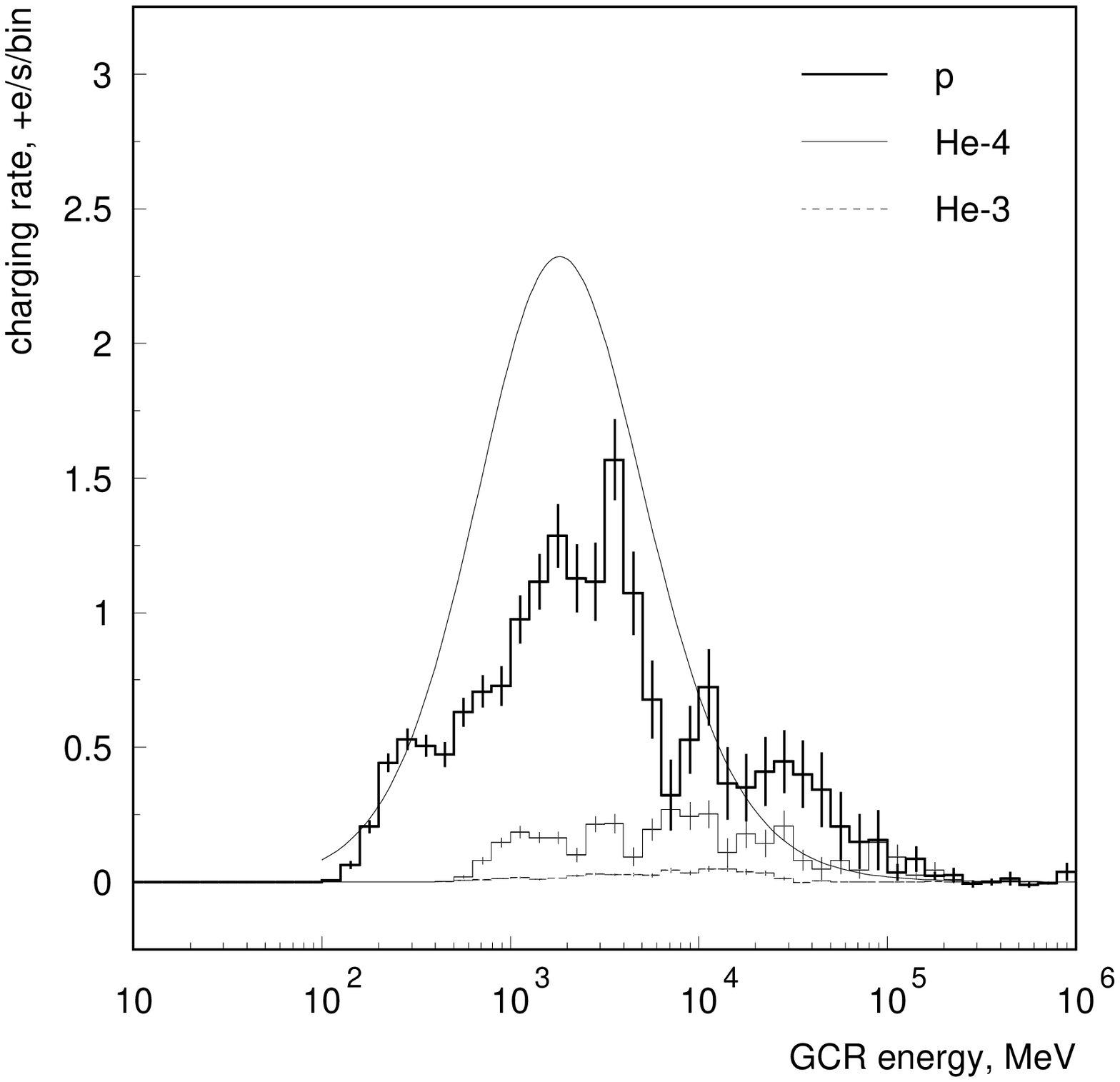,width=7cm}}
    \caption{Charging rate of TM$_0$ as a function of primary energy
    at solar minimum (left) and solar maximum (right); the smooth
    lines represent the GCR proton spectra (frequency per histogram
    bin, on different scales).}
   \label{spectrum}
\end{figure}

Cosmic rays contribute to the overall charging rate according to their
energy in the way shown in Fig.~\ref{spectrum}; also plotted are the
GCR proton spectra (now in relative flux per histogram bin, not per
unit energy). To a first approximation, the charging spectrum follows
the respective input spectrum above $\sim$100~MeV/n for protons as
well as He nuclei, and a factor $\simeq$2 ratio of solar minimum to
solar maximum rates reflects approximately the ratio between the GCR
fluxes.  Yet, the two scenarios have distinct spectral features: apart
from the shift towards higher energies caused by the harder solar
maximum flux, a peak is visible at lower energies in the solar minimum
situation.

\begin{figure}[ht]
   \centerline{\epsfig{file=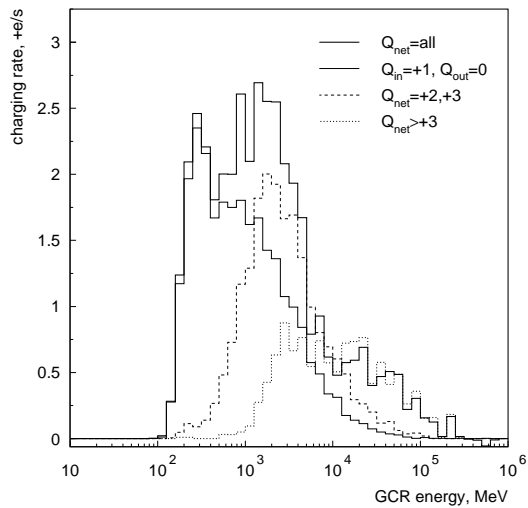,width=7cm}}
   \caption{Analysis of the proton charging rate for TM$_0$.}
    \label{spectrum2}
\end{figure}

Fig.~\ref{spectrum2} decomposes the effect of solar minimum protons
according to the actual net charge stopping and being ejected from the
TM. The feature at a few hundred MeV is readily explained in terms of
single positive charges stopping in the TM, which are mostly
protons. A mass overburden of at least 10--15~g/cm$^2$ encloses the
test masses, which shields against primaries below 100~MeV/n. The test
cube itself represents 90~g/cm$^2$ --- it takes a $\sim$400~MeV proton
to get through it. The rate of 100--400~MeV protons directly stopping
in the TM is approximately 10~+e/s at solar minimum (25\% of the
proton-induced rate) and negligible for the harder solar maximum
spectrum.

\begin{figure}[ht]
   \centerline{\epsfig{file=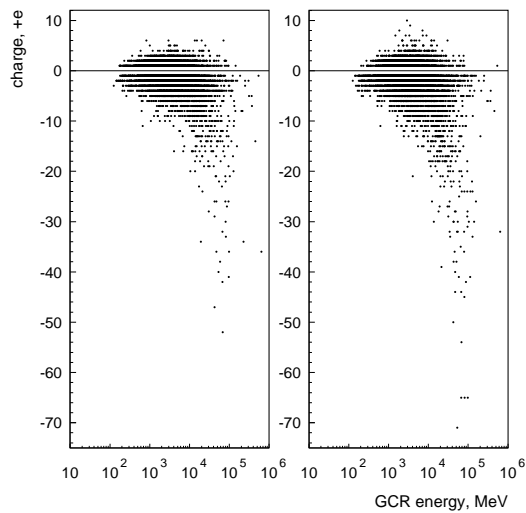,width=7cm}}
   \caption{Distribution of charges entering (left) and leaving
    (right) TM$_0$ as a function of primary energy for solar minimum
    protons.}
   \label{specinout}
\end{figure}

Increasingly energetic primaries result in higher net charges, as
e$^+$/e$^-$ pair production and cascaded nuclear reactions take
over. This is illustrated in Fig.~\ref{specinout}, which shows the
charges stopping and being ejected from the TM as a function of
primary energy --- only for those events producing a net
result. High-energy events are quite complicated, occasionally leading
to $\sim$100 negative particles (mostly electrons) stopping in the TM
while comparable numbers are ejected from it. In spite of some
cancellation, the net charging is positive. Secondary electrons are
responsible for a charging rate of +28~+e/s for solar minimum protons
(almost 70~\% of the proton-induced rate).

Although electromagnetic showers can create many e$^+$/e$^-$ pairs and
gamma rays, nuclear reactions are the most important charging
mechanism at higher energies, effectively multiplying the number of
primaries in the spacecraft. This multiplication role of hadronic
interactions was confirmed by a simple simulation carried out with a
basic geometry of concentric spheres providing 13~g/cm$^2$ of
shielding to one cubic TM. The charging rate induced by solar minimum
protons suffering electromagnetic processes alone was almost 3 times
smaller than that obtained when hadronic processes were added. This
fact partly explains the higher rates obtained in the present work.

\begin{figure}[ht]
  \centerline{\epsfig{file=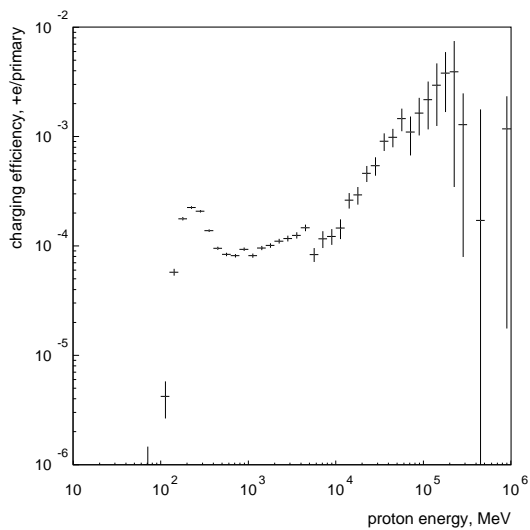,width=7cm}}
   \caption{Spectral charging efficiency for protons.}
   \label{efficiency}
\end{figure}

By dividing the charging spectrum by the input flux one obtains a
spectral charging efficiency, $Q(E_0)/N_0(E_0)$, shown in
Fig.~\ref{efficiency} for protons. The charging threshold at 100~MeV
is clearly seen in the figure, as is the initial peak previously
explained in terms of protons stopping in the TM. In this region, and
up to approximately 10~GeV, the charging efficiency is 10$^{-4}$
positive charges for each proton randomly emitted from the generator
sphere. The discontinuity visible near 6~GeV is artificially created
by the change-over between inelastic scattering models for protons
(intra-nuclear cascade and quark-gluon string models). At very high
energies the charging efficiency decreases to zero (within errors).

\section{Solar Energetic Particles}
\label{SEP}

\subsection{Characterisation of Solar Particle Events}

Irregular fluxes of solar particles (mainly protons) will add to the
steady background of GCRs, leading to further charging of the LISA
test masses. Solar events associated with radio bursts or moderate
x-ray flares are relatively short-lived ($\sim$hours) and very
frequent ($\sim$1000/year). However, their energy spectrum is usually
too soft ($<$50~MeV) to penetrate the shielding overlying the TMs. A
second kind of phenomena, known as solar energetic particle (SEP)
events, are associated with shock acceleration in coronal mass
ejections and their spectrum can extend up to hundreds of
MeV. Although much rarer than solar flares, SEP events can have large
particle fluences above 100~MeV and deposit significant amounts of
charge over periods of days. It is important to characterise these
events in terms of their frequency of occurrence, duration, peak flux
and total fluence as well as energy spectrum in order to calculate
their effects.

\begin{figure}[ht]
  \centerline{
    \epsfig{file=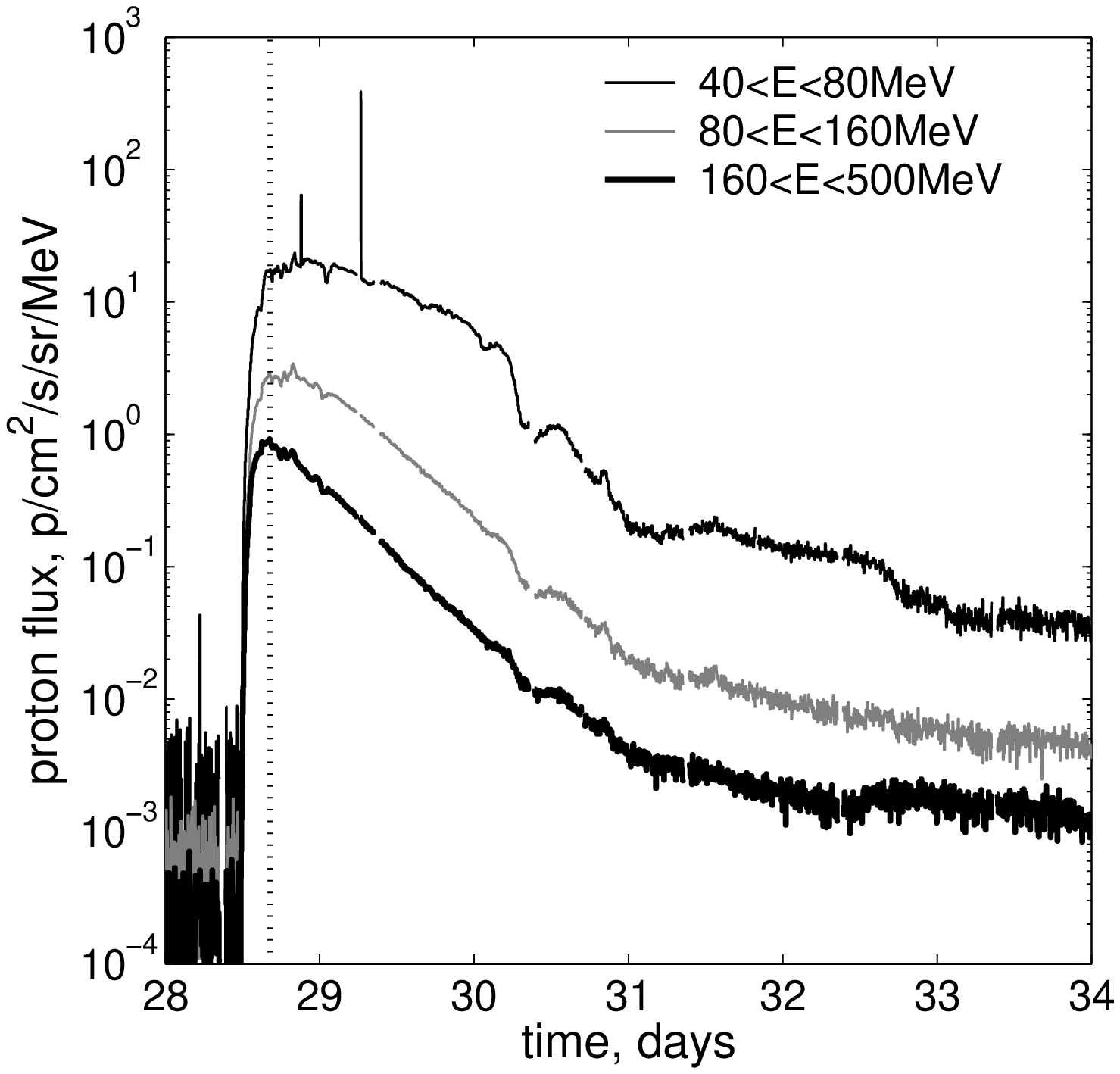,width=7cm}
    \epsfig{file=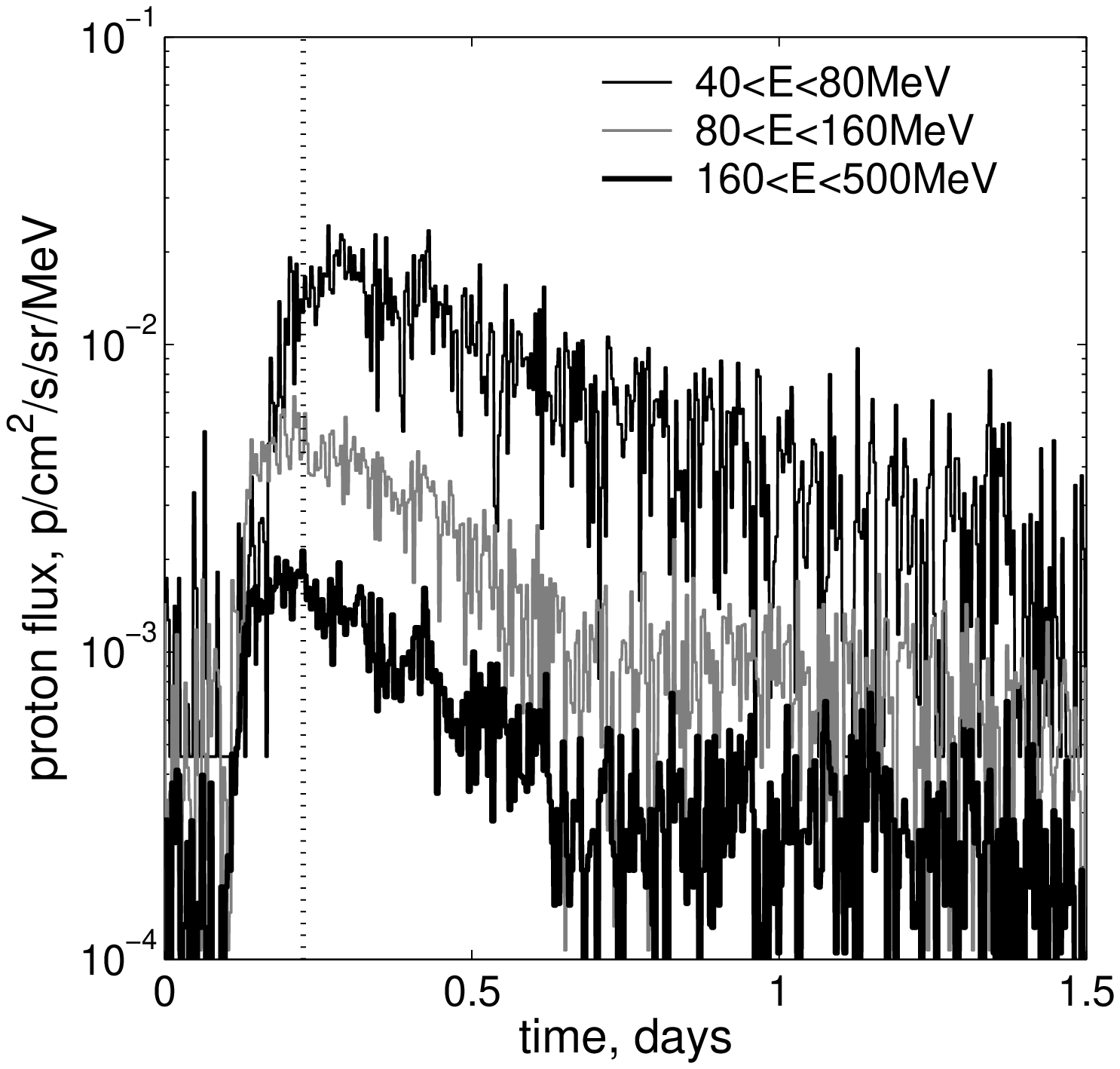,width=7cm}}
   \caption{Timelines of proton fluxes in three energy ranges measured
   by GOES satellites for Event~1 (29 Sept 1989) and Event~2 (20 May
   2001). Note that the fluxes preceding the two events are higher
   than expected for solar maximum. The time at which the peak flux in
   the highest energy band is reached is also shown.}
   \label{goes}
\end{figure}

We have investigated the charging potential of two SEP events which
can have operational implications for the LISA mission. Event 1
occurred on 29 Sept 1989 and is a well-known extreme event, with a
fluence $\Phi_{40}\!\simeq\!10^9$~protons/cm$^2$ above 40~MeV, which
caused widespread instrument malfunction in space (and on
Earth!). Event~2 is a small, unremarkable event registered on 20 May
2001, with a fluence of $\Phi_{40}\!\simeq\!6\times
10^5$~protons/cm$^2$. The respective proton fluxes recorded by the
Geostationary Operational Environmental Satellites (GOES) \cite{goes}
are shown in Fig.~\ref{goes}.

\begin{figure}[ht]
  \centerline{\epsfig{file=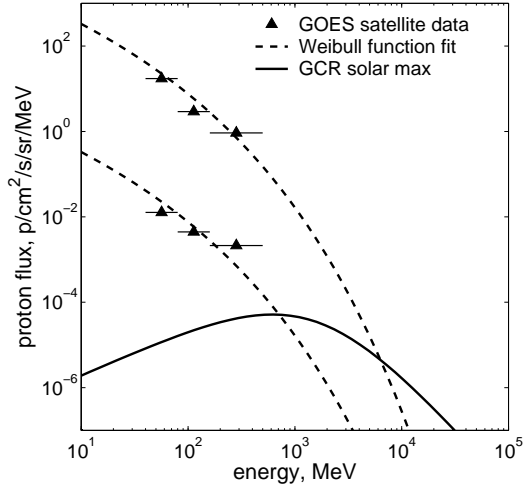,width=7cm}}
   \caption{Proposed differential energy spectra at peak flux for
   Event~1 and Event~2 using a Weibull function fit and the GOES
   data. Also shown is the spectrum for GCR protons at solar maximum.}
   \label{SEPspectra}
\end{figure}

The Weibull function is found to describe many SEP energy spectra. The
Weibull differential flux (in protons/cm$^2$/s/sr/MeV), is:
\begin{equation}
   \phi(E)= A\,k\,\alpha\,E^{\alpha-1} \exp{(-k E ^{\alpha})},
   \label{weibull}
\end{equation}
where $E$ is the primary energy and $A$, $k$ and $\alpha$ are free
parameters. Event~1 has been fitted with such a function using data
from space-borne detectors and ground-level neutron monitors together
with radiation transport codes \cite{dyer03}. This fit is shown in
Fig.~\ref{SEPspectra}. The peak fluxes in three GOES channels are also
plotted in the figure.

Events with small fluence are usually not amenable to a similar
parameterisation. Nevertheless, Event~2 was fitted with a Weibull
function with the same spectral indices ($\alpha$ and $k$) found for
Event~1 but with $A$ scaled to agree with the GOES data. This approach
should be sufficient for the estimation of charging rates. The result
is also shown in Fig.~\ref{SEPspectra}. Note that this spectrum
excludes the GCR background, which should be added in charging
calculations.

Finally, the SEP event rate must be established from the relative
distribution of event sizes together with the predicted rates for the
occurrence of a particular event fluence. The well-known Nymmik model,
based on a statistical analysis of historical records, applies to
event fluences $\Phi_{30}$ ($E\!>\!30$~MeV) in the range
10$^{5}$--10$^{11}$ protons/cm$^2$ \cite{nymmik99}. In this model, the
mean occurrence frequency, $\left<N(t)\right>$ (events/year), of SEP
events with fluence $\Phi_{30}\!>\!10^5$~protons/cm$^2$ is a function of
solar activity through the average monthly sunspot number, $W$:
\begin{equation}
   \left<N(t)\right> = 0.3\,W(t)^{0.75}.
   \label{SEPfrequency}
\end{equation}
In the same study, the distribution of event fluences is fitted with a
power law:
\begin{equation}
   \frac{1}{\left<N\right>}\frac{dN(\Phi_{30})}{d\Phi_{30}} 
   \propto \Phi_{30}^{-1.41}.
   \label{SEPsize}
\end{equation}

\begin{figure}[ht]
  \centerline{
    \epsfig{file=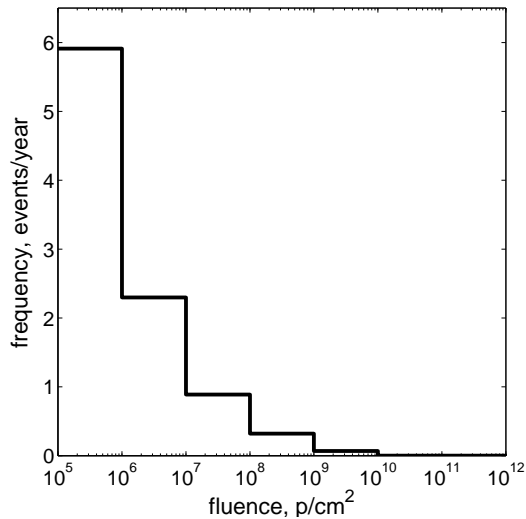,width=7cm}}
    \caption{SEP event frequency distribution as a function of event
     fluence for an average sunspot number $W$=100.}
   \label{SEPdistribution}
\end{figure}

\begin{figure}[ht]
  \centerline{
    \epsfig{file=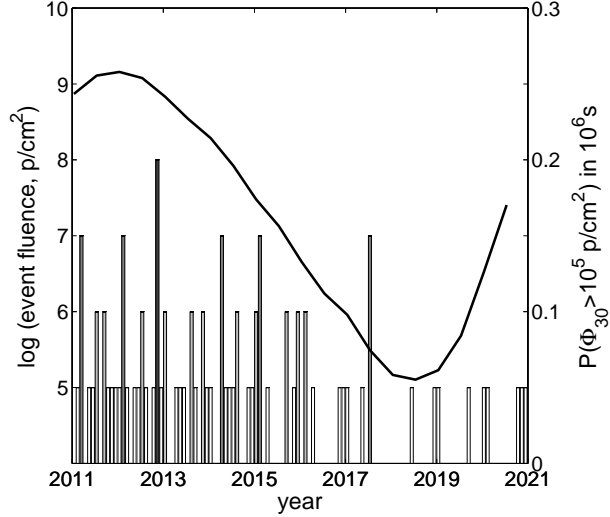,height=7cm,clip=}}
   \caption{Histogram: Predicted scenario of SEP events for the
   lifetime of the LISA mission. Line: Probability for a fluence
   $\Phi_{30}>10^{5}$~protons/cm$^2$ to occur during a 10$^6$~s period
   along the solar cycle.}
   \label{SEPprojection}
\end{figure}

These equations allow us to estimate the SEP event rate shown in
Fig.~\ref{SEPdistribution}. A sunspot number $W$=100 was considered to
be representative of the solar maximum conditions prevailing around
2012 when LISA is scheduled for launch. This analysis confirms that
events as large as Event~1 occur at most once in the 11-year solar
cycle (note that $W$ decreases either side of solar maximum). Smaller
ones such as Event~2 should occur some 6 times per year. The total
number expected near solar maximum is of order 10. In
Fig.~\ref{SEPprojection} we present a projection of events expected
during the mission lifetime. The same figure plots the probability for
a SEP fluence $\Phi_{30}>10^{5}$~protons/cm$^2$ to occur in a 10$^6$~s
period, assuming a Poisson distribution with mean given by
Eq.~\ref{SEPfrequency} ($W(t)$ is calculated from the average of 13
solar cycles). Such period represents a hypothetical extension of the
LISA bandwidth presently under discussion. This calculation suggests
that such a long dataset can easily be contaminated by a solar event.

\subsection{Charging Rates from SEP Events}

\begin{figure}[ht]
  \centerline{\epsfig{file=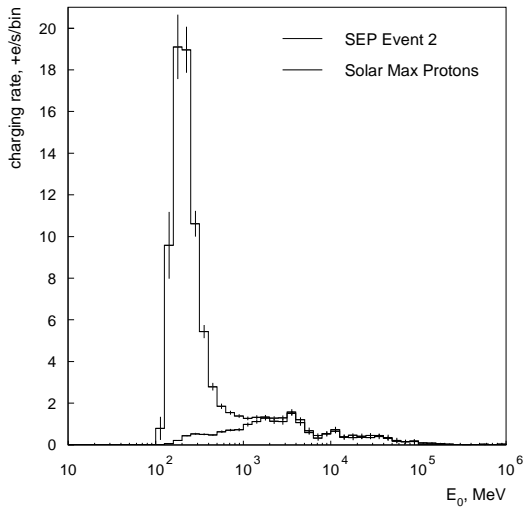,width=7cm}}
   \caption{Charging spectra for small SEP event and GCR protons at
   solar maximum.}
   \label{SEPcharging}
\end{figure}

The charging expected from these SEP events is obtained by multiplying
the differential energy spectra by the spectral charging efficiency
previously shown in Fig.~\ref{efficiency}. The charging spectrum for
Event~2 added to solar maximum protons is shown in
Fig.~\ref{SEPcharging}. The charging rate at peak flux is estimated at
87$\pm$3~+e/s (the error reflects only the MC uncertainty). This is
some 4 times above the solar maximum rate. On the other hand, Event~1
could charge the LISA test masses at $\sim$70~000~+e/s at peak flux!
Considering a simple exponential decay of the fluxes shown in
Fig.~\ref{goes}, we obtain the total charge deposits indicated in
Table~\ref{table2}.

\input{table2.tex}

This analysis is reassuring in that the number of SEP events with the
potential to produce spurious signals is relatively small even at the
peak of solar activity. This situation can change in periods of
exceptional solar turbulence ($W$=200 was observed in the most intense
solar cycle on record). The occurrence of solar events should be
correlated with the science data. Proton monitors aboard the LISA
spacecraft could be a very useful diagnostic tool for this
purpose. Earth-based monitors (or those on other missions) may miss
spatial effects only visible on the LISA constellation, or even on one
of its spacecraft alone.

\section{Non-simulated Physics}
\label{nonsim}

The MC simulation was complemented by an assessment of other physics
processes for their potential for TM charging. We categorise these
into kinetic emission from surfaces, photon emitting processes and
isotope production in the IS. The starting point for this analysis is
the MC calculation of the primary and secondary particle fluxes
tallied at the surface of the TMs. The latter are henceforth
considered to be made entirely from gold.

\subsection{Kinetic Emission Processes}

Low energy electrons ($\sim$eV) are emitted from surfaces bombarded by
electrons and ions. For fast (but non-relativistic) projectiles this
phenomenon is often known as kinetic emission, usually abbreviated as
EIEE or IIEE for electron- and ion-induced electron emission,
respectively. EIEE is exploited, for example, in scanning electron
microscopy (SEM) and electron multipliers. Useful reviews are
available on this subject \cite{springer1,springer2,icru} --- although
somewhat outdated in terms of experimental data. The analysis
presented here includes data more relevant to the case of gold
surfaces and high primary energies. Normal incidence is assumed for
simplicity.

\begin{figure}[ht]
   \centerline{\epsfig{file=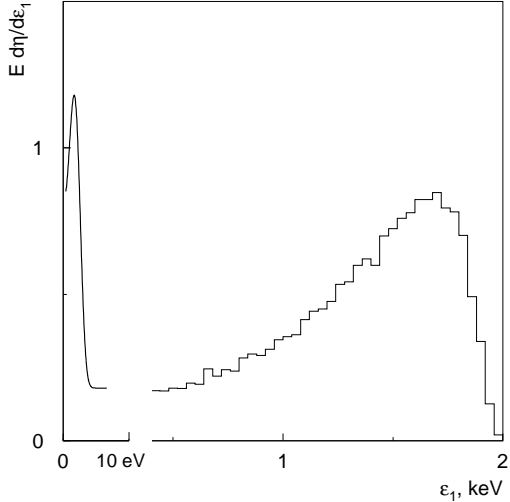,width=7cm}}
   \caption{Illustration of the electron spectrum emitted from a gold
   surface irradiated by 2~keV electrons at normal incidence. Thin
   line: backscattered primaries (Geant4 simulation); Thick line:
   kinetic electron emission.}
   \label{kinetic1}
\end{figure}

In EIEE, the energy spectrum of electrons emitted from the solid
extends from close to the incident electron energy, $E$, down to less
than 1~eV, and has the generic shape illustrated in
Fig.~\ref{kinetic1}. The higher-energy part of the spectrum is due to
backscattered primaries, the total number of which is characterised by
a reflection yield, $\eta$. This process is reproduced by Geant4. A
significant electron population is also represented in the figure at
energies below $\sim$10 eV. (Note that the total yield can be greater
than 1) These are secondary electrons created by the incident particle
which find their way to the solid surface. An electron is emitted if
its kinetic energy perpendicular to the boundary is greater than the
surface barrier (for metals, this corresponds to the work function,
$\phi$). This secondary electron yield, $\delta$, has a very soft
spectrum, peaking at a few eV. This peak is really due to the opposing
trends of the delta-ray production spectrum and the work function
subtraction at the surface, and has no particular meaning beyond
that. Naturally, the secondary electron yield is not taken into
account in the simulation, and has to be assessed separately.

Fast ions, such as H and He nuclei, also lead to electron emission
(IIEE). In this instance the backscattering probability is small, but
the kinetic emission process is otherwise very similar to EIEE. An
estimate of this contribution to TM charging is also required.

In addition, one must consider secondary emission from transmitted
electrons and ions, i.e. those ejected forward from the electrodes and
from the TMs by primaries travelling from the solid into vacuum.
Accurate experimental information on these forward yields is scarce,
but there is indication that they are similar to the backward yields
\cite{reimer77}.

Electrons emitted by EIEE and IIEE are generated extremely close to
the surface of the solid, in a layer $\lesssim$50~\AA~ thick. This effect
is therefore very sensitive to the surface conditions. Oxide layers on
metals and semiconductors are expected to increase the yield since
insulators usually have a higher emission yield than metals, although
different metals were observed to change in different directions when
exposed to oxygen \cite{benka98}. No change was observed for Au
targets in the same study. Nevertheless, the possible adsorption of
foreign species at the surface, namely organic deposits, should be
considered.

\subsubsection{EIEE and IIEE yields}

According to the theory of Schou~\cite{schou80}, which explores the
similarity between kinetic emission of electrons and atom sputtering,
the secondary yield caused by an incident non-relativistic electron
with energy $E$ is given by:
\begin{equation}
   \delta_{PE} = \beta{^\star} S_e(E) \Lambda \, ,
   \label{delta1}
\end{equation}
where $S_e(E)$ is the electronic stopping power at the primary
electron energy $E$, the parameter $\Lambda$ depends only on the
target material and $\beta{^\star}$ is a dimensionless function almost
insensitive to energy. This proportionality of the secondary electron
yield to the electronic stopping power has long been recognised. This
treatment ignores the contribution of secondaries generated by
backscattered electrons, which must be accounted for \cite{dubus87}:
\begin{equation}
   \delta = \delta_{PE}(1+\beta\eta).
   \label{delta2}
\end{equation}
In this expression, $\beta$ is the ratio of the mean secondary
electron generation of one backscattered electron to that of one
primary electron and $\eta$ is the backscattering coefficient.

\begin{figure}[ht]
   \centerline{
     \epsfig{file=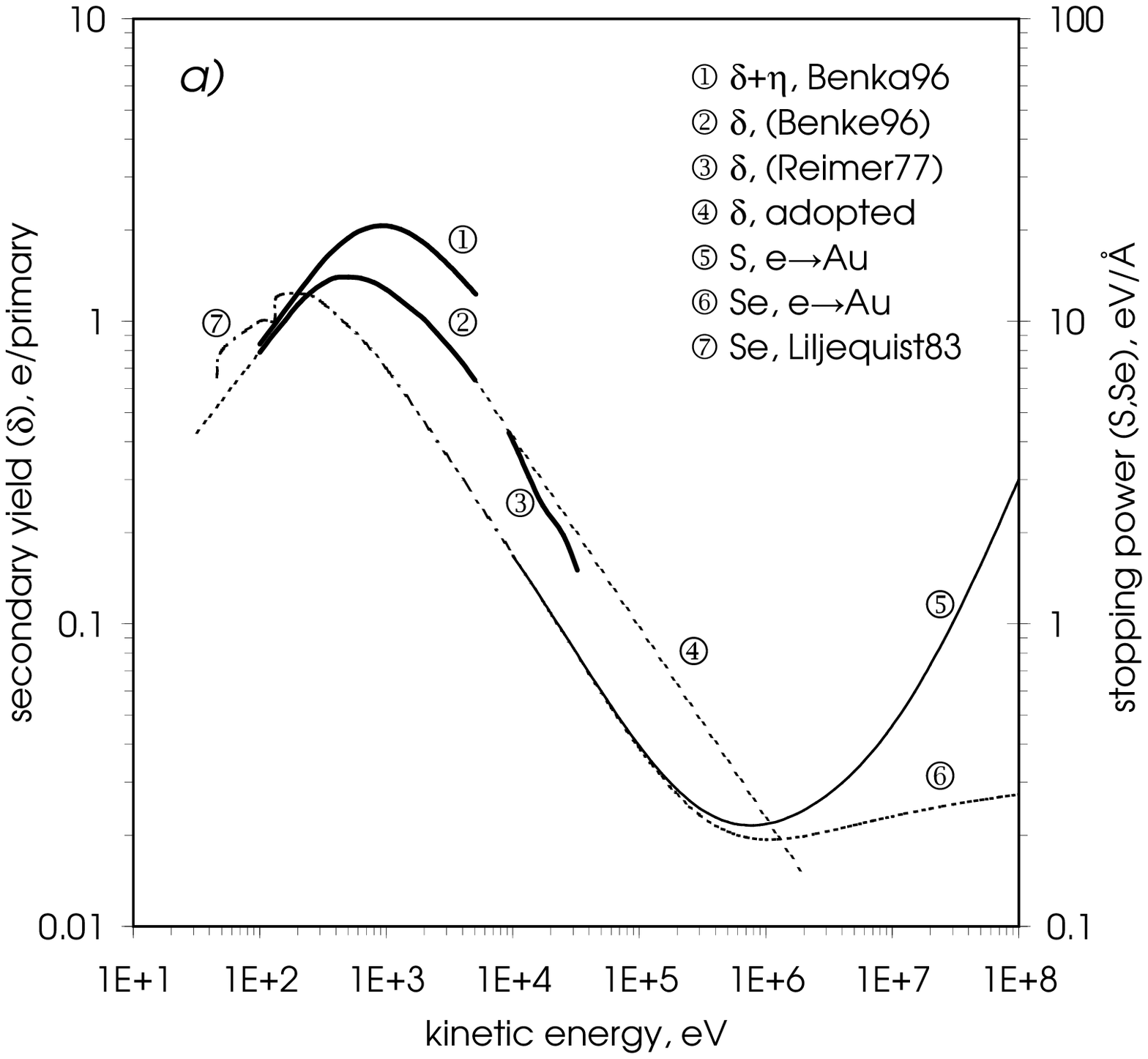,width=7cm}
     \epsfig{file=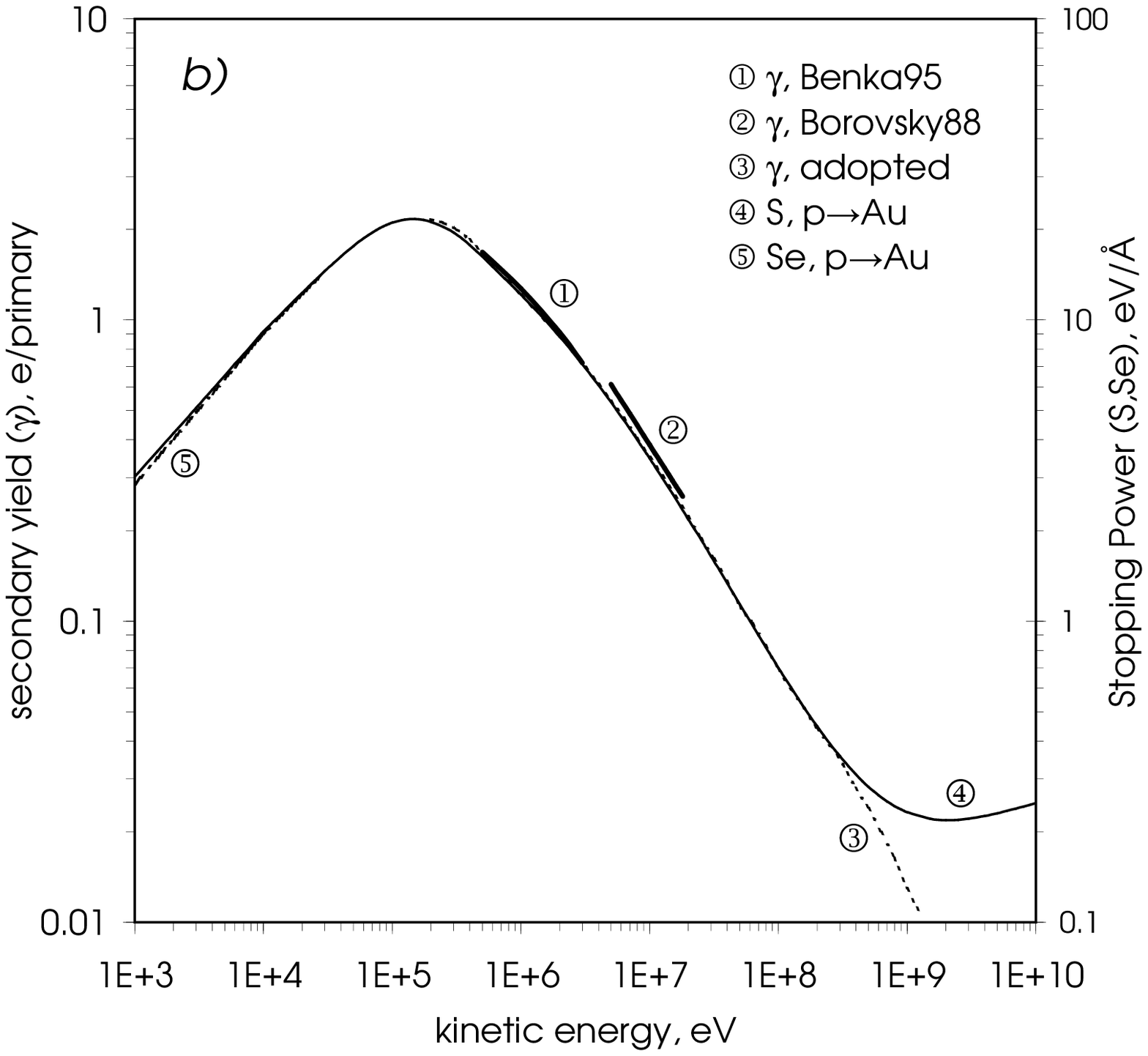,width=7cm}}
   \centerline{
     \epsfig{file=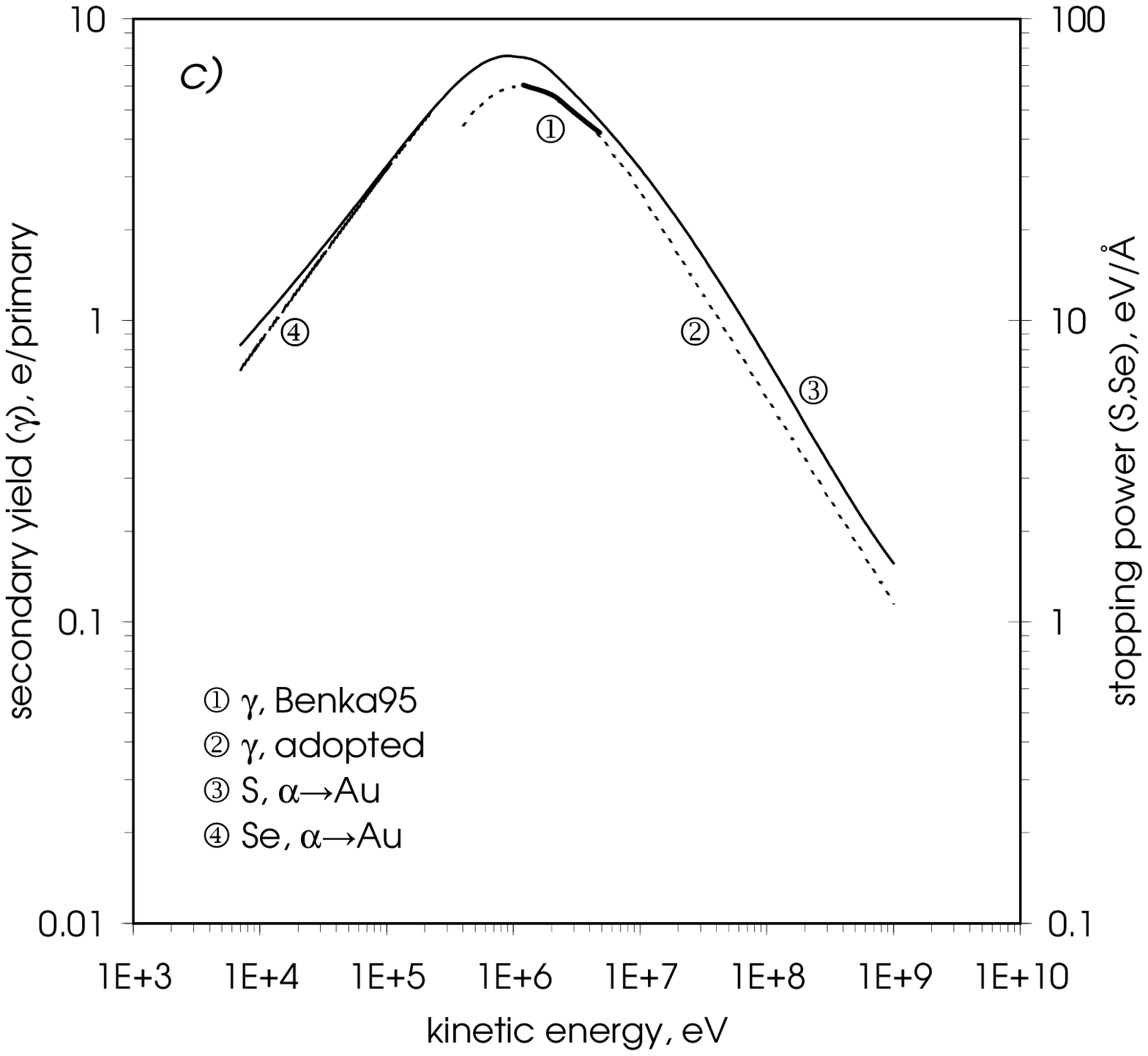,width=7cm}
     \epsfig{file=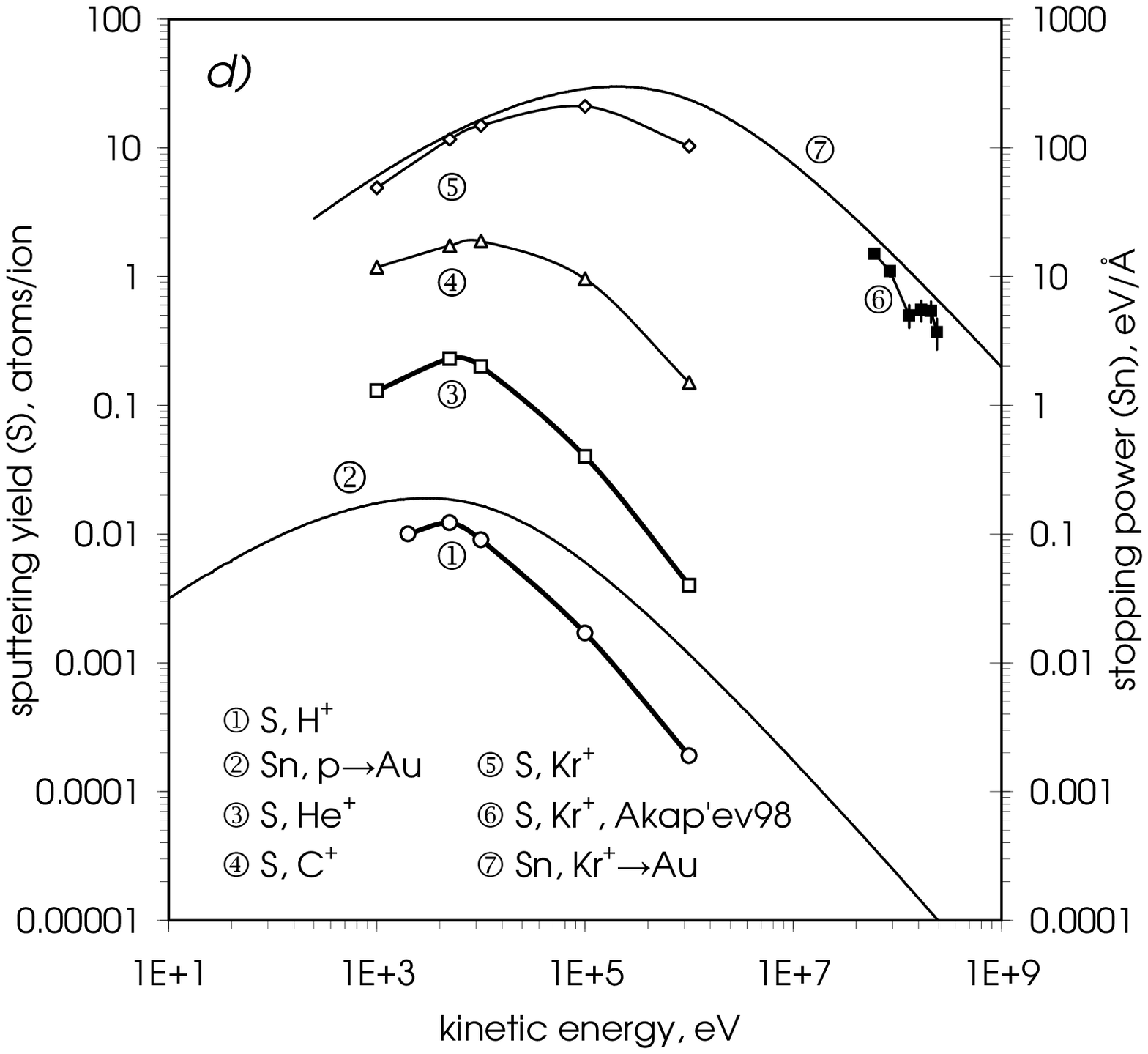,width=7cm}}
   \caption{a) EIEE yield and stopping power for electrons in gold.
   b) IIEE yield for protons. c) IIEE yield for alpha particles. d)
   Sputtering yield and nuclear stopping powers for several
   projectiles in gold.}
   \label{kinetic2}
\end{figure}

The parameters $\delta_{PE}$ and $\beta$ in Eq.~\ref{delta2} have been
measured for thin gold films for electron energies in the range
10--100~keV \cite{reimer77}. In addition, we consider $\eta$=0.5 in
this energy range, in order to add the contribution of secondaries
produced by backscattered primaries to the experimental values.
Experimental data were also found for lower energies \cite{benka96},
which reveal a total electron yield peaking at $\simeq$2 near
1~keV. These data include reflected primaries, which are a sizeable
fraction even at 1~keV. Backscattering has already been modelled by
Geant4, and must therefore be subtracted. The two datasets are shown
in Fig.~\ref{kinetic2}~a), along with the stopping power for electrons
in gold over a wide energy range \cite{star}. Curves 2 and 3 show our
corrected estimates for $\delta$. There is fair agreement between the
energy dependence of the two datasets and that of the stopping power
above $\sim$1~keV (the abrupt behaviour of $S_e$ at $\sim$100~eV is
due to the particular model considered \cite{liljequist83}). We have
adopted a secondary electron yield, $\delta$, which attempts to
reproduce the experimental data whilst following the energy dependence
of the stopping power in the non-relativistic Bethe-Block region
(10~keV--1~MeV). This is indicated by curve 4 in
Fig.~\ref{kinetic2}~a).

The theory of ion-induced emission is simpler than that of EIEE since
primary backscattering is not significant. The ion-induced emission
yield, $\gamma$, of secondary electrons by fast, non relativistic ions
is approximately proportional to the electronic component of the
stopping power for the ion in the material. Proton yields are
comparable to those of electrons of similar speed. To obtain a
charging rate estimate we make a similar extrapolation of low energy
experimental measurements to the relevant part of the primary
projectile spectrum. Note that this dependence is being extrapolated
to relativistic energies, which are beyond the applicability of the
theory.

Fig.~\ref{kinetic2}~b) shows the adopted yields for protons based on
data found in Ref.~\cite{benka95} and Ref.~\cite{borovsky88}. Here too
the energy dependence of the yield is in good agreement with that of
the stopping power over the energy range of the
measurements. Multiply-charged ions have larger yields, reflecting
higher rates of energy loss in the material. At GCR energies the
helium flux is composed of He$^{2+}$ rather than singly-ionised
atoms. Fig.~\ref{kinetic2}~c) shows the adopted yield based on the
data from Ref.~\cite{benka95}. Note that $\delta\simeq 6$ at the peak
of the stopping power.

\subsubsection{Kinetic emission rates}

\begin{figure}[ht]
   \centerline{\epsfig{file=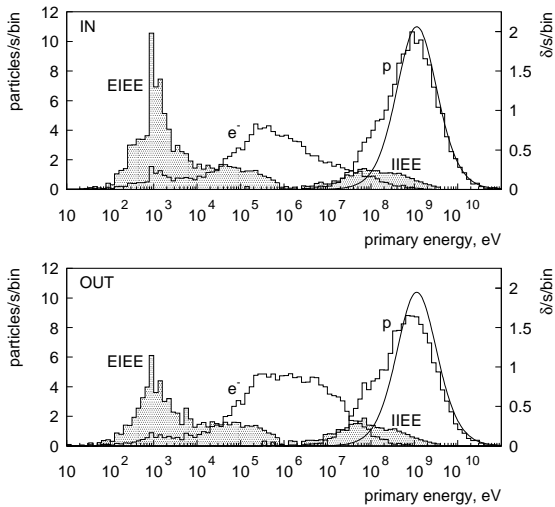,width=7.5cm}}
   \caption{Spectrum of electrons and protons entering (top) and
   leaving (bottom) one TM for incident protons at solar minimum and
   respective kinetic emission rates (shaded).}
   \label{kinetic3}
\end{figure}

The energy spectra of electrons and protons tallied at the TM surface
--- regardless of whether or not they lead to a net event charge ---
are shown in Fig.~\ref{kinetic3}. Similar data were generated for
solar maximum protons as well as $^4$He primaries. It is worth
pointing out the noticeable discontinuity observed in the electron
spectrum near 800~eV, which is due to the limitation of the
proton-induced delta-ray production model in Geant4 mentioned in
Section~\ref{physics}. A large uncertainty should therefore be
attributed to the sub-keV part of the energy spectra, as it is
difficult to estimate the number of electrons suppressed by the cut
(although no major error in the charging rate is expected).

\input{table3.tex}

On the same plot we show the corresponding emission rates obtained by
multiplying the primary spectra by the normal incidence yields. The
net rates for primaries entering and leaving the TMs are summarised in
Table~\ref{table3}. If identical yields are considered for both
incident and transmitted particles (forward and backward emission),
and also for the TM as well as the electrode surfaces, then these four
contributions should nearly cancel out statistically. However, this
cancellation can be compromised by sensing and actuating voltages
applied to the electrodes. Even if the net rates do cancel out, the
stochastic nature of the emission will be a source of spurious
acceleration noise in the sensor. For this reason, Table~\ref{table3}
also indicates the {\em average} emission rates
($\smash{^1/_2}(\delta+\gamma)_{\rm IN} +
\smash{^1/_2}(\delta+\gamma)_{\rm OUT})$, which we take to represent
the importance of this effect (and should not be added to $R$ as a net
charging rate). The average rate for solar minimum is 30~+e/s, a
figure comparable with the MC rate and larger than He-induced
charging.

\subsubsection{Atom sputtering}

Sputtering is the ejection of atoms from a surface under bombardment
from energetic ions. Although similar in nature to kinetic electron
emission, this is mainly a momentum-transfer process between
nuclei. Consequently, it depends on the nuclear stopping power, $Sn$,
rather than on the electronic component. Although sputtered species
are likely to be electrically neutral (although this is not always the
case) charge emission may occur from surfaces struck by the sputtered
atoms.

The sputtering yields of gold surfaces --- the number of atoms ejected
per incident ion --- were obtained with the SRIM2003 package
\cite{srim2003}, a software widely used for calculation of ion
interactions in matter. SRIM reproduces the experimental yields found
in Ref.~\cite{florio87} for bombardment of Au with singly-ionised
noble gases at 1-100 keV energies. The sputtering yields, $S$, for H,
He, C, Fe and Kr ions were then calculated, using a full-cascade
simulation, at ion kinetic energies between 1 keV and 1 MeV. These are
shown in Fig.~\ref{kinetic2}~d). Carbon is the third most abundant GCR
species, after H and He.

The nuclear stopping powers for protons and krypton ions in gold are
also shown in the figure. Experimental data were found for krypton
above $\sim$100 MeV \cite{akapev98}. For this projectile, an
extrapolation of $S$ to high energies based on $Sn$ seems
justified. Adopting a similar trend for the lighter projectiles allows
us to conclude that sputtering rates are negligible, much smaller than
1 atom/s. They can therefore be safely ignored.

\subsection{Photon Emission Processes}

We now examine non-simulated photon processes which can contribute to
TM charging. In this category we include x-ray transition radiation
(XTR), x-ray Cherenkov radiation (XCR) and particle-induced x-ray
emission (PIXE). Their effects may be two-fold: electron emission can
be due to self-absorption within the bulk material or photoemission
from a surface across the sensor gaps.

A surface irradiated by x-rays emits a primary spectrum of fast
photoelectrons and Auger electrons as well as low energy (eV)
secondaries generated by kinetic emission. The quantum efficiency (QE)
of gold surfaces can be as high as 10\% for photon energies in the
range 20 eV--10 keV \cite{day81}. Secondary electron emission
dominates the total QE in most materials. For gold, the primary
spectrum constitutes only 3--20\% of the total yield over that energy
range \cite{henke81}. A rapid decrease in efficiency is observed into
the UV region, as the photon energy approaches the surface work
function. An electron yield of $\sim$10$^{-5}$ electrons per absorbed
photon is expected at 5~eV~\cite{krolikowski70}. The low UV QE abates
our concern with sources of optical emission, such as optical
transition radiation (OTR) and optical Cherenkov emission (OCR).

\subsubsection{Proton-induced x-ray emission}

The ionisation cross-section of a particular atomic shell increases
with projectile energy in the MeV range, peaking when the projectile
velocity equals the electron velocity in that particular shell. A
simple rule-of-thumb for the peak energy is $E_0=1840AU_i$, where $A$
is the projectile mass number and $U_i$ is the electron binding energy
\cite{johansson88}. In gold, this corresponds to proton energies of
approximately 150 MeV for the K shell ($U_i$=80 keV) and 25~MeV for
the L shell (14~keV). The K shell fluorescence yield, $\omega_K$, is
practically unity for Au (whereas $\omega_L \approx 0.35$), which
means that nearly all K-shell ionisation ends up in x-ray emission
rather than Auger electrons. Given the large primary energies
involved, one must confirm that photon fluxes across the sensor gaps
are small. Although a prototype PIXE model does exists in Geant4, it
still reproduces experimental data poorly in the present version of
the toolkit. Instead, we have estimated the emission yields of K and L
x-rays for protons incident on a thick gold sample using the procedure
described in Ref.~\cite{johansson88}.

For a proton of energy $E_0$ entering a sample at angle $\theta_i$, the
PIXE yield, $Y$, radiated into a solid angle $d\Omega$ around an x-ray
exit angle $\theta_o$ is given by:
\begin{equation}
   Y(E_0,\theta_i,\theta_o) = \frac{N_A \omega_i f_\alpha}{A}
        \frac{d\Omega}{4\pi}
	\int_{E_0}^{0}\frac{\sigma(E)T(E)}{S(E)}dE .
   \label{pixe1}
\end{equation}
In this expression, $N_A$ is Avogadro's number, $\omega_i$ is the
fluorescence yield for the $i$-th atomic shell, $f_\alpha$ is the
fraction of fluorescence for a particular x-ray (K$_{\alpha 1},
K_{\alpha 2}$, etc.), $A$ is the molar mass of the material,
$\sigma(E)$ is the ionisation cross-section for the shell in question,
$S(E)$ is the proton stopping power and $T(E)$ describes the x-ray
self-absorption in the sample:
\begin{equation}
   T(E) = \exp\left(-\int_{E_0}^{E} \mu
          \frac{\cos(\theta_i)}{\cos(\theta_o)}\frac{1}{S(E)}\right) ,
   \label{pixe2}
\end{equation}
where $\mu$ is the mass attenuation coefficient at the x-ray
energy. The integrals in Eq.~\ref{pixe1} and Eq.~\ref{pixe2} must be
computed numerically. The proton stopping power is approximated by a
power law in the 1--1000 MeV range. For the K- and L-shell ionisation
cross-sections, the parameterisation proposed in
Ref.~\cite{johansson76} is used. Although intended for lower energies,
the cross-sections predicted for Au agree satisfactorily with
experimental data for K-shell ionisation by 160~MeV protons
\cite{jarvis72} and L x-ray production at 30~MeV \cite{shafroth73}.

\begin{figure}[ht]
   \centerline{\epsfig{file=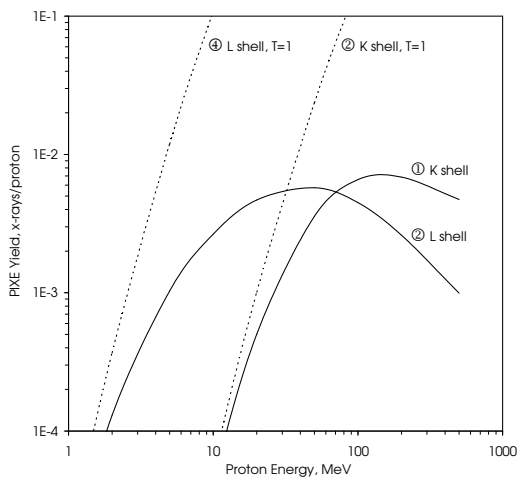,width=7cm}}
   \caption{K- and L-shell PIXE yields from protons on a thick Au
   target (normal incidence). The dotted lines describe the yields
   uncorrected for self-absorption ($T(E)$=1).}
   \label{pixe}
\end{figure}

The PIXE yields into the vacuum hemisphere are shown in
Fig.~\ref{pixe} for normal incidence protons. This estimate indicates
a value just under 0.01~photons/proton over the energy range of
interest. Considering maximum QEs of 0.1 for L- and K x-rays, the
proton rates given in Table~\ref{table3} result in a charging rate
below 1~+e/s. The photoelectron emission rate from x-ray
self-absorption in the sample itself is even smaller.

\subsubsection{Transition radiation}

Transition radiation (TR) consists of the emission of photons by fast
charged particles crossing a boundary between two media with different
dielectric constants. Optical photons are generated by slow, non- or
slightly-relativistic incident particles (OTR), whereas
ultra-relativistic projectiles produce preferentially x-rays (XTR).
Although common radiators are dielectrics, any material boundary can
generate TR photons, including metal/vacuum interfaces. In the latter
case, large photon yields are not foreseen due to the opaque nature of
the radiator. OTR is not expected to be an important source of TM
charging due to both the very small photon absorption length and the
relatively low photoelectric coefficients of gold at optical/UV
wavelengths. X-ray emission should be estimated in view of the large
flux of relativistic particles across the TM boundaries.

A particle with charge $ze$ crossing from vacuum into a material with
plasma frequency $\omega_p$ will produce a TR spectrum which diverges
at low energies and decreases rapidly for photon energies above
$\gamma\hbar\omega_p$, where $\gamma$ is the Lorentz factor. The
number of photons generated above an energy $\hbar\omega_0$ is given
by the following equation \cite{jackson}:
\begin{equation}
   N \simeq \frac{\alpha z^2}{\pi}
   \left[ \left( \ln\frac{\gamma \hbar\omega_p}{\hbar\omega_0}-1\right)^2
   + \frac{\pi^2}{12} \right] ,
   \label{xtr}
\end{equation}
where $\alpha$ is the fine structure constant. As this equation
suggests, the number of photons emitted above a fixed fraction of
$\gamma\hbar\omega_p$ is constant. For a gold radiator ($\hbar\omega_p
\simeq 8.5$~eV) the quantum yield above 0.1$\gamma\hbar\omega_p$
($\simeq$10~eV) for 10~GeV incident protons is equal to 0.59\%; 10~MeV
electrons generate the same yield above 18~eV TR energy. These small
values are further attenuated by the opacity of gold at such low
photon energies and the small QE of the gold surfaces opposite. No
significant charging is thus expected from TR photons crossing the
sensor gaps.

On the other hand, one can consider that most TR photons are absorbed
in the solid and create photoelectrons very close to the surface, a
fraction of which can be ejected --- possibly accompanied by kinetic
emission. Considering the TR and EIEE yields involved, together with
the simulated particle fluxes across the TM boundary, we place an
upper limit of 1~+e/s for the charging contribution from transition
radiation.

\subsubsection{X-ray Cherenkov radiation}

Cherenkov radiation (CR) is closely related to TR but, whereas the
latter occurs at discontinuities of the dielectric constant, Cherenkov
photons are emitted when the particle velocity is greater than the
speed of light in the bulk medium. Optical Cherenkov emission (OCR) in
transparent media is not important here for the reasons stated above
for OTR. At higher energies, comparable to atomic levels, CR is
absorbed over very short distances and so it manifests itself as
ionisation created along the particle track. This is already included
in the MC as a continuous energy loss term~\cite{grichine03}.

X-ray Cherenkov radiation (XCR) is suppressed in the limit of a
transparent medium since photons emitted from different parts of the
particle track interfere destructively, but this interference is
broken by a finite photon attenuation length, $l$. The ceramic slabs
supporting the electrodes, shown in Fig.~\ref{LISAInertialSensor}, can
act as XCR radiators. As in XTR, self-absorption near the surface is
probably more important than the x-ray flux across the IS gaps.

We have estimated the number of Cherenkov x-rays emitted from the
ceramic electrode supports into the forward hemisphere for 1~GeV
protons using the following yield equation \cite{bazylev82}:
\begin{equation}
   \frac{\d^3 N}{\hbar\d\omega \d x \d\theta^2} 
   = \frac{\alpha}{\pi\hbar c} \frac{\omega}{c} \theta^2\, \Im(Z),
   \label{xcr1}
\end{equation}
where $\hbar\omega$ is the photon energy, $x$ is the trajectory
length, $\theta$ is the photon emission angle with respect to the
particle trajectory and $Z$ is known as the complex formation zone of
the medium: $Z=L/(1-iL/l)$, with:
\begin{equation}
   L = \frac{c}{\omega}
   \left[ \gamma^{-2} + \frac{\omega_p^2}{\omega^2} + \theta^2 \right]^{-1}.
   \label{xcr2}
\end{equation}
Note that a transparent medium ($l$=$\infty$) implies $Z$=$L$, which
eliminates the yield in Eq.~\ref{xcr1}. For the ceramics, the CR yield
from a layer with thickness equal to $l$$\sim$0.1~mm is
0.025~photons/proton. Although only a fraction of the photoelectrons
created over a thickness $l$ escapes the electrode surface, this
process seems capable of TM charging at the level of 1~+e/s.

\subsection{Cosmogenic Activation}

The mechanisms analysed so far involve TM hits which happen, for all
practical purposes, in coincidence with the primary cosmic ray. We now
assess whether the activation of materials by GCR protons can lead to
significant TM charging through the subsequent radioactive decay of
relatively short-lived radioisotopes. The main concern is activation
of the TM itself, since the IS housing provides enough shielding from
external radioactive products.

The activity of a particular daughter species after an irradiation
time $t$ is \cite{krane}:
\begin{equation}
   {\mathcal A}(t) = 0.6 \frac{m}{A}\, \sigma\, \Phi (1-e^{-\lambda t}) 
   \quad {\rm [Bq] \, ,}
   \label{activation}
\end{equation}
where $m/A$ is the dimensionless ratio of the total mass to atomic
mass of the material, $\sigma$ is the cross-section in barn, $\Phi$ is
the proton flux in p/s/cm$^2$ and $\lambda$ is the decay constant of
the radioisotope. If the irradiation period is long compared to the
half-life of a particular radioisotope, then the single-species
activity in a 46-mm TM is given by ${\mathcal A}=6\sigma\Phi$.

Radioisotope production by proton irradiation has been studied in the
100--2000~MeV energy range for many targets \cite{michel97}. For gold,
the production cross-sections are highest for daughter isotopes just
below $^{179}$Au (spallation) and decrease quickly for lighter
elements. Overall, a dozen or so radioisotopes are produced at
$\sim$0.1~barn per nuclide; these decay quickly (relative to the
mission duration) mainly by electron capture (EC). We estimate a
secular-equilibrium activity of the order of 2~Bq for each spallation
product. Only a very small fraction of these decays ($\sim$0.1\%),
which occur randomly in the TM volume, can actually eject charge from
its surface. Therefore, this contribution to the charging rate can be
safely ignored. Radioisotope production by neutron capture was also
found to be similarly ineffective in TM charging.

\section{Conclusions}

The MC simulation of the interaction of GCR protons and He nuclei with
the LISA spacecraft indicates TM charging rates of nearly 25~+e/s for
solar maximum conditions, rising to 50~+e/s for solar
minimum. Table~\ref{table4} summarises the main results. We consider
an additional error of $\pm$30\% on the charging rates to account for
uncertainties in the GCR spectra, physics models and geometry
implementation. Although most of the charging is inflicted by protons,
He contributes disproportionately to its abundance due to the double
ion charge.

\input{table4.tex}

The solar minimum rate is almost twice that previously obtained with
Geant4 \cite{araujo03}, which is explained by the far more detailed
geometry implementation and the use of more complex physics models.
The new result is almost 5 times larger than the original Geant3
simulation \cite{jafry97}. Apart from the smaller 40~mm TM considered
in that study, the difference is also explained by insufficient
hadronic models in Geant3. A comparison of the respective charging
spectra reveals that most charging in that study resulted from direct
primary stopping in the test mass.

Several potential charging mechanisms were assessed independently of
the simulation. Some x-ray processes were found to be able to
contribute to TM charging at a maximum level of $\pm$1~+e/s. More
significantly, the kinetic emission of low energy electrons due to
bombardment of the sensor by electrons and ions was identified as a
more serious concern. The net charging from kinetic emission will
depend on a number of factors, but even a perfect cancellation of the
emission across the IS gaps will not avoid some acceleration
noise. Although there are considerable uncertainties in these
estimates, it is clear that this effect can rival the MC rates, and
even eclipse the contributions of GCR nuclei beyond H. In this context
we believe that the experimental determination of IIEE yields at
relativistic GCR energies merits future attention.

Shot-noise fluctuations with a spectral density of 24~+e/s/Hz$^{1/2}$
are predicted at solar minimum (i.e. charge fluctuations of $24/(2\pi
f)$~+e/Hz$^{1/2}$). That figure increases to 28.4~+e/s/Hz$^{1/2}$ if
one includes the effect of kinetic emission and considers a continuous
charge neutralisation exactly balanced at a rate $-R$. A charging rate
of 400~+e/s in single charges would be required to produce this level
of fluctuations. The acceleration noise caused by charging could reach
$4\times 10^{-16}$~m/s$^2$/Hz$^{1/2}$ at $f$=0.1~mHz for typical IS
parameters \cite{shaul04b}, making charging disturbances one of the
dominant noise sources at low frequencies.

Charging from SEP events was also investigated. Some 10 events per
year are expected to affect the LISA science data at solar maximum,
which is not a critical scenario. Further work is under way to
consolidate these predictions and to qualify the need for independent
in-flight diagnostic tools such as particle monitors.

\section{Acknowledgements}

This work was supported by the European Space Agency through the
SEPTIMESS project under Contract No. 16339/02/NL/FM. The authors wish
to thank the CERN IT Division for facilitating access to the LSF
Services. Thanks are also due to the Geant4 Collaboration for their
effort in responding to our requirements. We thank S.~Merkowitz
(NASA/GSFC) for providing the LISA solid model, P.~Sarra (CGS) for the
LTP sensor model, N.~Dunbar (EADS Astrium UK) and H.~Stockburger (EADS
Astrium Germany) for the SMART-2 spacecraft model, and D.~Smart
(SSTD/RAL) for the caging mechanism design. Thanks are due to
C.~Grimani (U.~Urbino/INFN), H.~Vocca (U.Perugia/INFN), E.~Daly
(TOS-EES/ESTEC), F.~Lei (QinetiQ), G.Santin (TEC-EES/ESTEC) and
L.Desorgher (U.Bern) for their assistance and suggestions.


\end{document}

%% file: table1.tex
\begin{sidewaystable}
\centering
\begin{minipage}{22.0cm}
\centering
\caption{\label{table1} Summary of the Monte Carlo Results}
\vspace{2mm}
\begin{tabular}{cc|c|cc|ccc|ccc}
\hline
primary  & solar    & GCR flux          & \multicolumn{2}{c|}{timeline}   & \multicolumn{3}{c|}{TM$_0$}             & \multicolumn{3}{c}{TM$_1$}                      \\
particle & activity & $F$, /s/cm$^2$   & $N_0, 10^8$  & $T$, s & $R$, +e/s& $\sigma_M$, e/s& $S_R$, e/s/Hz$^{1/2}$& $R$, +e/s& $\sigma_M$, e/s&$S_R$, e/s/Hz$^{1/2}$\\
\hline
p        &          & 4.29              & 121.1       & 400    & 39.8     & 0.8             & 21.2                  & 41.2     & 0.8             & 21.5                \\
$^4$He   & min      & 0.315             & 14.4        & 642    &  7.2     & 0.3             & 10.5                  &  7.7     & 0.3             & 11.0                \\
$^3$He   &          & 0.0591            & 14.1        & 3366   & 1.08     & 0.05            & 3.95                  & 1.04     & 0.05            & 4.05                \\
\hline
total    &          & 4.66              & 149.6       & --     & 48.1     & 0.9             & 24.0                  & 49.9     & 0.9             & 24.5                \\
\hline
p        &          & 1.89              & 53.3        & 400    & 17.8     & 0.6             & 16.8                  & 19.7     & 0.6             & 17.9                \\
$^4$He   & max      & 0.142             & 9.3         & 924    & 3.6      & 0.2             & 8.79                  &  3.5     & 0.2             & 9.06                \\
$^3$He   &          & 0.0236            & 8.0         & 4804   & 0.45     & 0.03            & 2.77                  & 0.44     & 0.03            & 2.9                 \\
\hline
total    &          & 2.06              & 70.6        & --     & 21.8     & 0.6             & 19.2                  & 23.7     & 0.6             & 20.3                \\
\hline
\end{tabular}
\end{minipage}
\end{sidewaystable}

%% file: table2.tex
\begin{table}
\centering
\begin{minipage}{\linewidth}
\centering
\caption{Charging from SEP Events}
\label{table2}
\vspace{2mm}
\begin{tabular}{lcc}
\hline
                        & Event 1        & Event 2        \\
                        & 29 Sept 1989   & 20 May 2001    \\
\hline
$\Phi_{30}$, protons/cm$^2$   & $1\times 10^9$ & $6\times 10^5$ \\
event duration, days    & $\sim$5        & $\sim$1        \\
decay time, days        & 0.43           & 0.34           \\
rate at peak flux, +e/s & 68,000         & 87             \\
charging from SEP, +e   & $3\times 10^9$ & $3\times 10^6$ \\
charging from GCR, +e   & $1\times 10^7$ & $2\times 10^6$ \\
\hline
\end{tabular}
\end{minipage}
\end{table}

%% file: table3.tex
\begin{sidewaystable}
\centering
\begin{minipage}{20.0cm}
\centering
\caption{\label{table3} Estimates of EIEE and IIEE rates.}
\vspace{2mm}
\begin{tabular}{cc|cc|cc|cc|cc|c}
\hline
primary  & solar    & \multicolumn{2}{c|}{electrons in} & \multicolumn{2}{c|}{electrons out} & \multicolumn{2}{c|}{hadrons in} & \multicolumn{2}{c|}{hadrons out} & $\langle$K.E.$\rangle \dag$ \\
particle & activity & e/s        & $\delta$/s        & e/s           & $\delta$/s & h/s        & $\gamma$/s & h/s         & $\gamma$/s & +e/s  \\
\hline
p        &          & 112        & 17.6       & 141           & 14.0       & 154        & 4.7        & 145        & 5.5       & 20.9         \\
$^4$He   & min      & 24         & 2.3        & 31            & 2.2        & 19         & 5.7        & 21         & 8.0       & 9.1          \\
$^3$He   &          & --         & --         & --            & --         & --         & --         & --         & --        & --           \\
\hline
total    &          & 136        & 19.9       & 172           & 16.2       & 173        & 10.4       & 166        & 13.5      & 30.0         \\
\hline
p        &          & 72         & 12.0       & 94            & 8.7        & 80         & 2.1        & 84         & 3.0       & 12.9         \\
$^4$He   & max      & 14         & 1.2        & 19            & 1.3        & 10         & 3.4        & 17         & 4.7       & 5.3          \\
$^3$He   &          & --         & --         & --            & --         & --         & --         & --         & --        & --           \\
\hline
total    &          & 86         & 13.2       & 113           & 10.0       & 90         & 5.5        & 101        & 7.7       & 18.2         \\
\hline
\end{tabular}
\end{minipage}
\footnotetext{$\dag$ Average kinetic emission from the TM and the inner electrode housing.}
\end{sidewaystable}

%% file: table4.tex
\begin{table}
\centering
\begin{minipage}{\linewidth}
\centering
\caption{Summary of Test-Mass Charging in LISA}
\label{table4}
\vspace{2mm}
\begin{tabular}{cc|cc|cc|c}
\hline
primary  & solar    & \multicolumn{2}{c|}{charging rate}            & \multicolumn{2}{c|}{charging fluctuations} & $R_{\!e\!f\!f}$  \\
particle & activity & $R\pm\sigma_M$, +e/s &$\delta$+$\gamma$, +e/s $\dag$ & $S_R$, e/s/Hz$^{1/2}$ & $S_{T}$ $\ddag$    &  e/s        \\
\hline
p        &          & 41.2$\pm$0.8         & 21                    & 21.4                  & --                 & --  \\
$^4$He   & min      &  7.7$\pm$0.3         &  9                    & 10.7                  & --                 & --  \\
$^3$He   &          & 1.04$\pm$0.05        & --                    & 4.0                   & --                 & --  \\
\hline
total    &          & 49.9$\pm$0.9         & 30                    & 24.2                  & 28.4               & 403 \\
\hline
p        &          & 19.7$\pm$0.6         & 13                    & 17.4                  & --                 & --  \\
$^4$He   & max      &  3.5$\pm$0.2         &  5                    &  8.9                  & --                 & --  \\
$^3$He   &          & 0.44$\pm$0.03        & --                    &  2.8                  & --                 & --  \\
\hline
total    &          & 23.7$\pm$0.6         & 18                    & 19.7                  & 22.5               & 254 \\
\hline
\end{tabular}
\footnotetext{$\dag$ Average kinetic emission as given in Table~3.}
\footnotetext{$\ddag$ Including $\delta$+$\gamma$ and continuous
discharge at a rate $-R$.}
\end{minipage}
\end{table}

%% file: HAraujo.bbl
\begin{thebibliography}{00}

\bibitem{prephaseA}
K.Danzmann {\it et al.}, LISA Pre-Phase A Report (2nd Ed.) MPQ~233 (1998).

\bibitem{vitale02}
S.Vitale {\it et al.}, Nucl. Phys. B 110 (1998) 209.

\bibitem{jafry96} 
Y.Jafry, T.J.Sumner, S.Buchman, Class. Quantum Grav. 13 (1996) A97.

\bibitem{jafry97}
Y.Jafry, T.J.Sumner, Class. Quantum Grav. 14 (1997) 1567.

\bibitem{sumner00}
T.J.Sumner, Y.Jafry, Adv. Space Res. 25(6) (2000) 1219.

\bibitem{araujo03}
H.M.Ara\'{u}jo {\it et al.}, Class. Quantum Grav. 20 (2003) S311.

\bibitem{shaul04a}
D.N.A.Shaul {\it et al.}, Class. Quantum Grav. 21 (2004) S647.

\bibitem{shaul04b}
D.N.A.Shaul {\it et al.}, to appear in Class. Quantum Grav. (2004).

\bibitem{geant3} 
{\it GEANT -- Detector Description and Simulation Tool}, 
CERN Program Library, Long Writeup W5013, 1993.

\bibitem{geant4}
Geant4 Collaboration, Nucl. Instrum. Meth. A 506 (2003) 250;
(release 06-00 was used in this work).

\bibitem{fluka} 
A. Fass\`{o} {\it et al.}, In: Workshop on Simulating Accelerator
Radiation Environments, Santa Fe, USA (1993).

\bibitem{vocca04} 
H.Vocca {\it et al.}, Class. Quantum Grav. 21(5) (2004) S665.

\bibitem{wass04}
P.Wass {\it et al.}, in preparation (2004).

\bibitem{LISASolidModel}
LISA Integrated Solid Model (NASA/GSFC) Rev 1, Sep 2003.

\bibitem{ISSolidModel}
LISA Inertial Sensor Design Report (Carlo Gavazzi Space) LTP-RP-CGS-001.

\bibitem{physicsman}
Geant4 Physics Reference Manual (http://geant4.web.cern.ch/geant4)

\bibitem{grimani03} 
C.Grimani {\it et al.}, Class. Quantum Grav. 21(5) (2004) S629.

\bibitem{catia}
C.Grimani, private communication.

\bibitem{LSF} 
Data produced at LSF Batch Services provided by CERN IT Division.

\bibitem{goes} 
The Geostationary Operational Environmental Satellite Program (GOES)
is a joint effort of NASA and the US National Oceanic and Atmospheric
Administration (NOAA) (http://spidr.ngdc.noaa.gov/spidr/).

\bibitem{dyer03}
C.S.Dyer {\it et al.}, IEEE Trans. Nucl. Sci. 50(6) (2003) 2038.

\bibitem{nymmik99}
R.A.Nymmik, Radiat. Meas. 30 (1999) 287.

\bibitem{springer1}
G.H\"{o}hler (Ed.), {\it Particle-Induced Electron Emission I},
(Springer, Vol. 122, 1991)

\bibitem{springer2}
G.H\"{o}hler (Ed.), {\it Particle-Induced Electron Emission II},
(Springer, Vol. 123, 1992)

\bibitem{icru}
Int. Commission on Radiation Units and Measurements (ICRU) Rep. 55, 1995.

\bibitem{reimer77}
L.Reimer, H.Drescher, J. Phys. D 10 (1977) 805.

\bibitem{benka98}
O.Benka {\it et al.}, Phys. Rev. A 58(4) (1998) 2978.

\bibitem{schou80}
J.Schou, Phys. Rev. B 22(5) (1980) 2141.

\bibitem{dubus87}
A.Dubus {\it et al.}, Phys. Rev. B 36(10) (1987) 5110.

\bibitem{benka96}
O.Benka {\it et al.}, Nucl. Instrum. Meth. B 117 (1996) 350.

\bibitem{star} 
Stopping Power and Ranges for Electrons (ESTAR), Protons (PSTAR) and
Alphas (ASTAR) (http://physics.nist.gov\-/PhysRefData\-/Star/\-Text).

\bibitem{liljequist83}
D.Liljequist, A. Appl. Phys. 16 (1983) 1567.

\bibitem{benka95}
O.Benka {\it et al.}, Phys. Rev. A 52(5) (1995) 3959.

\bibitem{borovsky88}
J.E.Borovsky {\it et al.}, Nucl. Instrum. Meth. B 30 (1988) 191.

\bibitem{srim2003} 
J.F.Ziegler, {\it SRIM2003: The Stopping and Range of Ions in Matter},\\
http://www.srim.org/\#SRIM.

\bibitem{florio87}
A.Oliva-Florio {\it et al.}, Phys. Rev. B 35(5) (1987) 2198.

\bibitem{akapev98}
G.I.Akap'ev {\it et al.}, Tech. Phys. 43(1) (1998) 120.

\bibitem{day81}
R.H.Day {\it et al.}, J. Appl. Phys. 52(11) (1981) 6965.

\bibitem{henke81}
B.L.Henke {\it et al.}, J. Appl. Phys. 52(3) (1981) 1509.

\bibitem{krolikowski70}
W.F.Krolikowski and W.E.Spicer, Phys. Rev. B 1(2), 487.

\bibitem{johansson88} 
S.A.E.Johansson and J.L.Campbell, {\it PIXE: A Novel Technique for
Elemental Analysis}, (Wiley, 1988).

\bibitem{johansson76}
S.A.E.Johansson and T.B.Johansson, Nucl. Instrum. Meth. 137 (1976) 473.

\bibitem{jarvis72}
O.N.Jarvis {\it et al.}, Phys. Rev. A 5(3) (1972) 198.

\bibitem{shafroth73}
S.M.Shafroth {\it et al.}, Phys. Rev. A 7(2) (1973) 556.

\bibitem{jackson}
J.D.Jackson, {\it Classical Electrodynamics}, 3rd Ed., (Wiley, 1998).

\bibitem{grichine03}
V.M.Grichine, Nucl. Instrum. Meth. A 502 (2003) 133.

\bibitem{bazylev82}
V.A.Bazylev and N.K.Zhevago, Sov. Phys. Usp. 25 (1982) 565.

\bibitem{krane}
K.Krane, {\it Introductory Nuclear Physics}, (Wiley, 1988) 465.

\bibitem{michel97}
R.Michel {\it et al.}, Nucl. Instrum. Meth. B 129 (1997) 153.

\end{thebibliography}
